\documentclass[graybox]{svmult}

\usepackage{mathptmx}       % selects Times Roman as basic font
\usepackage{helvet}         % selects Helvetica as sans-serif font
\usepackage{courier}        % selects Courier as typewriter font
\usepackage{type1cm}        % activate if the above 3 fonts are
\usepackage{makeidx}         % allows index generation
\usepackage{graphicx}        % standard LaTeX graphics tool
\usepackage{multicol}        % used for the two-column index
\usepackage[bottom]{footmisc}% places footnotes at page bottom

\makeindex            

\usepackage{array}
\newcolumntype{X}[1]{>{\raggedleft\arraybackslash}p{#1}}
\newcolumntype{P}[1]{>{\raggedright\arraybackslash}p{#1}}

\usepackage[fleqn]{amsmath}
\newtheorem{preremark}{Property}
\newenvironment{propertyNew}{\begin{preremark}\upshape}{\end{preremark}}

\newtheorem{preremark1}{Strategy}
\newenvironment{strategy}{\begin{preremark1}\upshape}{\end{preremark1}}

\usepackage{algorithm}
\usepackage{algorithmic}
\begin{document}
\bibliographystyle{spmpsci}

\title*{A comparative study of top-k high utility itemset mining methods}

%\titlerunning{Top-K high utility pattern mining}
\author{Srikumar Krishnamoorthy \\ Indian Institute of Management, Ahmedabad, India}
\authorrunning{Srikumar Krishnamoorthy}
\institute{Srikumar Krishnamoorthy \at Indian Institute of Management Ahmedabad, Gujarat, India \email{srikumark@iima.ac.in} \\ \\
This is a pre-print version of a book chapter. The final accepted version of this chapter can be referred in an upcoming Springer book titled "High-Utility Pattern Mining: Theory, Algorithms and Applications" Edited by Fournier-Viger P., Lin, C.-W., Nkambou, R., Vo. B., and Tseng, V.S.}

\maketitle

\abstract{High Utility Itemset (HUI) mining problem is one of the important problems in the data mining literature. The problem offers greater flexibility to a decision maker to incorporate her/his notion of utility into the pattern mining process. The problem, however, requires the decision maker to choose a minimum utility threshold value for discovering interesting patterns. This is quite challenging due to the disparate itemset characteristics and their utility distributions. In order to address this issue, Top-K High Utility Itemset (THUI) mining problem was introduced in the literature. THUI mining problem is primarily a variant of the HUI mining problem that allows a decision maker to specify the desired number of HUIs rather than the minimum utility threshold value. Several algorithms have been introduced in the literature to efficiently mine top-k HUIs. This paper systematically analyses the top-k HUI mining methods in the literature, describes the methods, and performs a comparative analysis. The data structures, threshold raising strategies, and pruning strategies adopted for efficient top-k HUI mining are also presented and analysed. Furthermore, the paper reviews several extensions of the top-k HUI mining problem such as data stream mining, sequential pattern mining and on-shelf utility mining.  The paper is likely to be useful for researchers to examine the key methods in top-k HUI mining, evaluate the gaps in literature, explore new research opportunities and enhance the state-of-the-art in high utility pattern mining.}

\section{Introduction}
\label{sec:1}
Frequent Itemset (FI) mining \cite{Agrawal1994Apriori,Han2000Fpgrowth,Zaki2000Eclat} is one of the most widely studied problems in the data mining literature. The problem involves determining the set of all itemsets whose co-occurrence frequencies are higher than the user specified frequency threshold. The generated frequent itemsets can be used to discover non-trivial and interesting patterns of customer behaviour. In retail business, a manager can use the discovered patterns to perform product assortment planning, determine pricing schemes, offer customized promotions and make effective shelf allocation decisions. Though the initial focus of the problem was on extracting interesting patterns from retail market basket data, the problem has wide variety of applications in numerous domains. For example, frequent itemset and rule mining have been successfully applied in areas like intrusion detection \cite{lee2000adaptive}, click-stream analysis \cite{mobasher2001effective}, review mining \cite{liu2005opinion}, e-commerce personalization \cite{lin2002efficient}, software error pattern analysis \cite{livshits2005dynamine} and business process mining \cite{djenouri2017extracting}.
\par 
High Utility Itemset (HUI) mining framework \cite{Liu2012HUIMiner,Liu2005TwoPhase,efimKBS2017} extends the basic frequent itemset mining framework and uses a notion of itemset utility. The utility of an item (or itemset) is a function of the internal utility (such as purchase quantity, count of clicks), and external utility (profit or margin of items). The generalized notion of utility in the new framework allows discovery of more interesting and actionable patterns (high utility itemsets) from databases. However, the high utility itemset does not satisfy downward closure property \cite{Liu2005TwoPhase}. This makes the problem considerably harder compared to a traditional frequent itemset mining problem. 
\par 
Several methods have been proposed in the literature to efficiently mine high utility itemsets. Most of these methods require specification of a minimum utility threshold value and differ primarily on the data structures, heuristics and pruning strategies used. The specification of a minimum utility threshold value, however, is a non-trivial task and require sufficient domain knowledge. For example, in a study conducted by Wu et al \cite{wu2012miningTKU}, the authors demonstrate that a small change in minimum utility threshold value (0.03\% to 0.02\% in the chainstore dataset) can result in significantly long execution times. In essence, improper choice of minimum utility threshold value can lead to one or more of the following issues: (1) generate very few (or zero) patterns, (2) generate too many patterns burdening the user with filtering of relevant patterns, and (3) incur significantly high computational overhead in generating large number of patterns. 
\par 
A trial and error process is commonly followed to determine a suitable minimum utility threshold value for a given dataset. This is a highly inefficient process. A decision maker is more interested in simple and intuitive queries of the form: ``What are the top-k interesting patterns that can be used for decision making?". Answering this query is a non-trivial task and has been the focus of the recent research in the literature \cite{duong2016efficientKHMC,tseng2016efficientTKO}. 
\par 
In the frequent itemset mining literature, several attempts have been made \cite{cheung2004mining,quang2006exminer,salam2012mining} to address the top-k frequent pattern mining problem. The top-k HUI mining problem is distinct from these methods and is proven to be more challenging \cite{duong2016efficientKHMC,tseng2016efficientTKO}. This paper aims to study the research trends in the area of top-k HUI mining and uncover research gaps that require further investigation.  
\par The rest of the paper is organized as follows: Section 2 describes the key definitions and notations used in top-K HUI mining. The top-k HUI mining problem is also formally stated in this section. Section 3 presents different approaches to top-k HUI mining. A detailed discussion on one-phase and two-phase top-k HUI mining methods are made in this section. Section 4 provides a comparative analysis of key top-k HUI mining methods in the literature. Subsequently, the top-k HUI mining variants such as stream mining, sequential pattern mining and on-shelf pattern mining methods are described. Section 6 presents open issues and future research opportunities in top-k HUI mining. Finally, Section 7 presents concluding remarks.

\section{Preliminaries and problem statement}
\label{sec:2}
This section presents the key definitions and notations and formally states the top-k HUI mining problem. The key definitions and notations used are as per the standard conventions followed in the top-k HUI mining literature \cite{duong2016efficientKHMC,Liu2012HUIMiner,tseng2016efficientTKO,wu2012miningTKU}.  
\par 
Let $I=\{i_{1}, i_{2} ... i_{m}\}$ be a set of distinct items. A set $X = \{x_{1}, x_{2} ... x_{p}\} \subseteq I, x_{i} \in I$ is referred as a p-itemset. 
\par 
A transaction $T_{j} = \{x_{l} | l=1,2...N_{j}, x_{l} \in I\}$, where $N_{j}$ is the number of items in transaction $T_{j}$. A transaction database $D$ has set of transactions and a sample transaction database is given in Table~\ref{table:SampleDB}.

\begin{table}[!h]
	\renewcommand{\arraystretch}{1.2}
	\caption{Sample transaction database}
	\label{table:SampleDB}
	%\centering
	\begin{tabular}{>{\centering}p{1.2cm} p{2.5cm} p{3cm} p{2.5cm} >{\centering}p{2cm}}
		\hline
		TID & Transaction & Purchase & Utility & TU \tabularnewline
		& & Qty (IU) & (U) & \tabularnewline 
		\hline
		$T_{1}$ & a, c, d, e, f &  1, 1, 1, 2, 2 & 5, 1, 2, 6, 2  & 16 \tabularnewline
		$T_{2}$ & a, c, e, g & 2, 6, 2, 5 & 10, 6, 6, 5 & 27 \tabularnewline 
		$T_{3}$ & a, b, c, d, e, f & 1, 2, 1, 6, 1, 5 & 5, 4, 1, 12, 3, 5 & 30 \tabularnewline
		$T_{4}$ & b, c, d, e &  4, 3, 3, 1 & 8, 3, 6, 3 & 20 \tabularnewline
		$T_{5}$ & b, c, e, g &  2, 2, 1, 2 & 4, 2, 3, 2 & 11 \tabularnewline 
		$T_{6}$ & a, c, d, e, f &  3, 3, 3, 3, 3 & 15, 3, 6, 9, 3 & 36 \tabularnewline 
		$T_{7}$ & a, b, c, d, f &  1, 1, 1, 2, 3 & 5, 2, 1, 4, 3 & 15 \tabularnewline 
		$T_{8}$ & a, b, c, e, f &  1, 2, 2, 1, 1 & 5, 4, 2, 3, 1 & 15 \tabularnewline 
		\hline
	\end{tabular}
\end{table}

\begin{table}[!h]
	\renewcommand{\arraystretch}{1.2}
	\caption{Item profits}
	\label{table:ItemProfits}
	%\centering
	%\begin{tabular}{cccccccc}
	\begin{tabular}{>{\centering}p{3.6cm} >{\centering}p{1cm} >{\centering}p{1cm} >{\centering}p{1cm} >{\centering}p{1cm} >{\centering}p{1cm} >{\centering}p{1cm} >{\centering}p{1cm}}
		\hline
		Item & a & b & c & d & e & f & g \tabularnewline 
		\hline
		Profit per unit in \$ (EU) & 5 & 2 & 1 & 2 & 3 & 1 & 1 \tabularnewline 
		\hline
	\end{tabular}
\end{table}

\begin{definition}\label{def:EU}
	Each item $x_{i} \in I$ is assigned an external utility value (e.g., profit), referred as $EU(x_{i})$. A sample set of items and their associated profit details are provided in Table~\ref{table:ItemProfits}. The item profits are assumed to be positive.       
\end{definition}

\begin{definition}
	Each item $x_{i} \in T_{j}$ is assigned an internal utility value (e.g., purchase quantity), referred as $IU(x_{i}, T_{j})$. For example, in Table~\ref{table:SampleDB}, $IU(e, T_{6}) = 3$.
\end{definition}

\begin{definition}
	The utility of an item $x_{i} \in T_{j}$, denoted as $U(x_{i}, T_{j})$, is a function of the internal and external utilities of items. It is computed as the product of external and internal utilities of items in the transaction, $T_{j}$. That is,
	\begin{equation}
	U(x_{i}, T_{j}) = EU(x_{i}) * IU(x_{i}, T_{j})
	\end{equation}
\end{definition}
For example, in Table~\ref{table:SampleDB}, $U(e, T_{6}) = EU(e) * IU(e, T_{6}) = 3 * 3 = 9$.

\begin{definition}\label{def:util}
	The utility of an itemset $X$ in transaction $T_{j}$ ($X \subseteq T_{j}$) is denoted as $U(X, T_{j})$. 
	\begin{equation}
	U(X, T_{j}) = \sum_{x_{i} \in X} U(x_{i}, T_{j})
	\end{equation}
\end{definition}
For example, in Table~\ref{table:SampleDB}, $U(\{acdef\}, T_{6}) = 15 + 3 + 6 + 9 + 3 = 36$.

\begin{definition}\label{def:miuItem}
	The minimum item utility of an item, denoted as $miu(x_{i})$, is defined as $MIN(U(x_{i}, T_{j}))$ where $1 \leq j \leq n, T_{j} \in D$, $x_{i} \in T_{j}$ and n is the total number of transactions in the database.
\end{definition}

\begin{definition}\label{def:support}
	The support of an item, denoted as $Sup(X)$, is defined as the count of transactions that contain the itemset $X$. For example, $Sup(\{ad\}) = 4$. 
\end{definition}

\begin{definition}\label{def:miu}
	The minimum utility of an itemset, denoted as $miu(X)$, is defined as $\sum_{x_{i} \in X} miu(x_{i}) * Sup(X)$. 
\end{definition}

\begin{definition}\label{def:mauItem}
	The maximum item utility of an item, denoted as $mau(x_{i})$, is defined as $MAX(U(x_{i}, T_{j}))$ where $1 \leq j \leq n, T_{j} \in D$ and $x_{i} \in T_{j}$.
\end{definition}

\begin{definition}\label{def:mau}
	The maximum utility of an itemset, denoted as $mau(X)$, is defined as $\sum_{x_{i} \in X} mau(x_{i}) * Sup(X)$. 
\end{definition}
The $miu$ and $mau$ of an itemset are used respectively as lower and upper bound utility value during the itemset mining process. 

For the sample database, the minimum and maximum item utility values are shown in Table~\ref{table:miuMAU}.

\begin{table}[!h]
	\renewcommand{\arraystretch}{1.2}
	\caption{Minimum and maximum item utility values}
	\label{table:miuMAU}
	%\centering
	\begin{tabular}{>{\centering}p{2.4cm} >{\centering}p{1.2cm} >{\centering}p{1.2cm} >{\centering}p{1.2cm} >{\centering}p{1.2cm} >{\centering}p{1.2cm} >{\centering}p{1.2cm} >{\centering}p{1.2cm}}
		\hline
		Item & a & b & c & d & e & f & g \tabularnewline 
		\hline
		miu & 5 & 2 & 1 & 2 & 3 & 1 & 2 \tabularnewline
		mau & 15 & 8 & 6 & 12 & 9 & 5 & 5 \tabularnewline 
		\hline
	\end{tabular}
\end{table}

\begin{definition}
	The utility of an itemset $X$ in database $D$ is denoted as $U(X)$.
	\begin{equation}
	U(X) = \sum_{X \subseteq T_{j} \in D} U(X, T_{j})
	\end{equation}
\end{definition}
For example, $U(\{acdef\}) = U(\{acdef\}, T_{1}) + U(\{acdef\}, T_{3}) + U(\{acdef\}, T_{6}) = 16 + 26 + 36 = 78$.

\begin{definition} \textit{Real Item Utilities (RIU)}
	The real item utility of 1-itemsets are denoted as $RIU = \{U(x_{1}), U(x_{2}), U(x_{3}) ... U(x_{m})\}$. Let the Kth highest utility value in $RIU$ be denoted as $RIU_{k}$.
\end{definition}
For the running example, the real item utility values are given in Table~\ref{table:riu}.

\begin{table}[!h]
	\renewcommand{\arraystretch}{1.2}
	\caption{Real item utilities}
	\label{table:riu}
	%\centering
	\begin{tabular}{>{\centering}p{2.4cm} >{\centering}p{1.2cm} >{\centering}p{1.2cm} >{\centering}p{1.2cm} >{\centering}p{1.2cm} >{\centering}p{1.2cm} >{\centering}p{1.2cm} >{\centering}p{1.2cm}}
		\hline
		Item, $x_{i}$ & a & b & c & d & e & f & g \tabularnewline 
		\hline
		$U(x_{i})$ & 45 & 22 & 19 & 30 & 33 & 14 & 7 \tabularnewline
		\hline
	\end{tabular}
\end{table}

\begin{definition}
	The transaction utility, $TU(T_{j})$ for transaction $T_{j}$ is defined as
	\begin{equation}
	TU(T_{j}) = \sum_{X \subseteq T_{j}  \hspace{0.3em} and  \hspace{0.3em} x_{i} \in X} U(x_{i}, T_{j})
	\end{equation} 
\end{definition}
For example, $TU(T_{2}) = U(a, T_{2}) + U(c, T_{2}) + U(e, T_{2}) + U(g, T_{2}) = 27$

\begin{definition}
	Let $T_{j}/X$ denote the set of all items after $X$ in $T_{j}$. For example, in Table~\ref{table:SampleDB}, $T_{1}/\{ad\} = {\{ef\}}$, $T_{2}/\{ae\} = {g}$. 
\end{definition}

\begin{definition}\label{def:rutil}
	The remaining utility of an itemset $X$ in transaction $T_{j} (X \subseteq T_{j})$, denoted as $RU(X, T_{j})$, is computed as,
	\begin{equation}
	RU(X, T_{j}) = \sum_{x_{i} \in (T_{j}/X)} U(x_{i}, T_{j}), \\
	\end{equation}
\end{definition}
For example, in Table~\ref{table:SampleDB}, $RU(\{ad\}, T_{1}) = 6 + 2 = 8$, $RU(\{ae\}, T_{2}) = 5$.

\begin{definition}
	The remaining utility of an itemset $X$ in database $D$ is denoted as RU(X).
	\begin{equation}
	RU(X) = \sum_{X \subseteq T_{j} \in D} RU(X, T_{j})
	\end{equation}
\end{definition}
For example, in Table~\ref{table:SampleDB}, $RU(\{ad\}) = RU(\{ad\}, T_{1}) + RU(\{ad\}, T_{3}) + RU(\{ad\}, T_{6}) +$ \\ \hspace*{14em} $RU(\{ad\}, T_{7}) = 8 + 8 + 12 + 3 = 31$.

\begin{definition}
	The absolute minimum utility value is denoted as $\delta$.  
\end{definition}

\begin{definition} (High Utility Itemset)
	An itemset $X$ is referred as a High Utility Itemset (HUI) iff its utility $U(X)$ is greater than or equal to the minimum utility threshold value $\delta$. 
\end{definition}

A high utility itemset is neither monotonic or anti-monotonic. That is, the utility of an itemset $U(X)$ is equal to, higher or lower than that of its supersets/subsets. For the sample database in Table~\ref{table:SampleDB}, the set of all high utility itemsets at $\delta = 59$ is given in Table~\ref{table:HUIs}.

\begin{table}[!h]
	\renewcommand{\arraystretch}{1.2}
	\caption{High utility itemsets}
	\label{table:HUIs}
	%\centering
	\begin{tabular}{>{\centering}p{1.2cm} p{2.5cm} >{\centering}p{1.5cm} >{\centering}p{1.2cm} p{2.5cm} >{\centering}p{1.5cm}}
		\hline
		S.No. & High Utility Itemsets & Utility & S.No. & High Utility Itemsets & Utility\tabularnewline
		\hline
		1 & a e c & 80 & 7 & a e & 67 \tabularnewline
		2 & f d a e c & 78 & 8 & f d a & 67 \tabularnewline
		3 & f d a c & 73 & 9 & d a e & 63 \tabularnewline
		4 & f d a e & 73 & 10 & f a e & 62 \tabularnewline
		5 & f a e c & 69 & 11 & d a c & 60 \tabularnewline
		6 & d a e c & 68 & 12 & a c & 59 \tabularnewline
		\hline
	\end{tabular}
\end{table}

\begin{definition} \label{def:twu}
	The transaction weighted utility of an itemset X, denoted as $TWU(X)$, is defined as
	\begin{equation}
	TWU(X) = \sum_{X \subseteq T_{j} \in D} TU(T_{j})
	\end{equation}
\end{definition}
For the transaction database in Table~\ref{table:SampleDB}, $TWU(g) = TU(T_{2}) + TU(T_{5}) = 27 + 11 = 38$. The TWU values for the sample transactional database in Table~\ref{table:SampleDB} is provided in Table~\ref{table:TWU}.

\begin{table}[!h]
	\renewcommand{\arraystretch}{1.2}
	\caption{Transaction weighted utility}
	\label{table:TWU}
	%\centering
	%\begin{tabular}{cccccccc}
	\begin{tabular}{>{\centering}p{2.5cm} >{\centering}p{1cm} >{\centering}p{1cm} >{\centering}p{1cm} >{\centering}p{1cm} >{\centering}p{1cm} >{\centering}p{1cm} >{\centering}p{1cm}}		
		\hline
		Item & g & b & f & d & a & e & c \tabularnewline
		\hline 
		TWU & 38 & 91 & 112 & 117 & 139 & 155 & 170 \tabularnewline
		\hline
	\end{tabular}
\end{table}

\begin{definition} (High Transaction Weighted Utility Itemset)
	An itemset $X$ is referred as a High Transaction Weighted Utility Itemset (HTWUI) iff its utility $TWU(X)$ is greater than or equal to the minimum utility threshold value $\delta$. 
\end{definition}

\begin{propertyNew} \textit{TWDC Property.} \label{def:twdc} If $TWU(X) < \delta$, then $\forall X' \supseteq X, \\ \hspace*{13em} TWU(X') \leq TWU(X) < \delta$. 
\end{propertyNew}

As per the apriori property, $Sup(X') \leq Sup(X)$. This implies that $TWU(X') \leq TWU(X) < \delta$. That is, the High Transaction Weighted Utility Itemset (HTWUI) satisfies the downward closure property. This property is commonly exploited as a key pruning strategy for mining high utility itemsets in the literature \cite{Ahmed2009IHUP,Liu2005TwoPhase}.

\begin{propertyNew} \textit{DGU: Discarding Globally Unpromising items property.} \label{property:dgu} If $TWU(x_{i}) < \delta$, then $x_{i}$ is an unpromising item and $\forall X \supseteq x_{i}, \hspace*{.1em} TWU(X) < \delta$. 
\end{propertyNew}

This property was introduced in \cite{Tseng2010UPGrowth} and is a sub-property of the TWDC property \cite{Ahmed2009IHUP,Liu2005TwoPhase}. The proof can be easily verified from the proof of the TWDC property.

\begin{definition} (Top-K High Utility Itemset)
	The set of all $K$ HUIs with the highest utilities in $D$ are denoted as $TopKHUI$.
\end{definition}

\begin{definition}
	The optimal minimum utility threshold value, denoted as $\delta_{F}$, is defined as
	\begin{equation}
	\delta_{F} = min\{U(X) | X \in TopKHUI\}
	\end{equation}
\end{definition}

\vspace{.3em}\noindent\textbf{Problem statement} Given a transactional database $D$ and the desired number of HUIs ($K$), the problem of top-k high utility mining involves determining $K$ HUIs in $D$ that have the highest utilities. 
\par 
It is possible for multiple HUIs to have the same utility value at the optimal $\delta_{F}$ value. For example, when K=3 (or 7), at the optimal value of $\delta_{F} =$ 73 (or 67) (refer to Table~\ref{table:HUIs}) there are multiple HUIs. Top-k HUI mining methods in the literature treat such boundary cases differently. Some works use a stricter value of K \cite{duong2016efficientKHMC} while others relax the value of K to extract all the HUIs at the optimal $\delta_{F}$ value \cite{tseng2016efficientTKO,wu2012miningTKU}. We follow the more stricter definition in the rest of the paper and apply it consistently across all our comparative evaluations. 
\par 
For the transaction database in Table~\ref{table:SampleDB}, when $K = 3$, $TopKHUI = \{\{aec\}: 80, \{fdaec\}: 78, \{fdac\}: 73\}$.

\section{Approaches to Top-K high utility itemset mining}
\label{sec:3}

The top-k HUI mining methods can be broadly categorized as one-phase and two-phase methods. Early approaches to top-k HUI mining relied on two-phases for mining top-k HUIs (TKU  \cite{wu2012miningTKU} and REPT \cite{ryang2015topREPT}). The two-phase methods first generate candidate top-k HUIs (phase 1) and then extract the relevant top-k HUIs (phase 2). One-phase methods, on the other hand, does not generate intermediate candidate top-k HUIs and directly mine the top-k HUIs. TKO \cite{tseng2016efficientTKO} and KHMC \cite{duong2016efficientKHMC} are the key one-phase top-k HUI mining methods proposed in the literature.  In this section, we review all of the top-k HUI mining methods in the literature. 

\subsection{Two-phase methods}
\label{subsec:m1}

The top-k HUI mining problem was first introduced by Wu et al \cite{wu2012miningTKU}. The authors discuss several challenges in adapting the top-k frequent pattern mining methods and the consequent need for design of new algorithms for top-k HUI mining. 

\subsubsection{TKU algorithm} \label{algo:tkuAlgorithm}

The TKU algorithm \cite{wu2012miningTKU} is one of the earliest algorithms for mining top-k HUIs from transactional databases. The algorithm mines the HUIs in two phases. In the first phase, the algorithm constructs the UP-Tree \cite{Tseng2010UPGrowth} and generates the Potential top-K HUIs (PKHUIs). Subsequently, the algorithm determines the top-k HUIs from the set of PKHUIs. Five different threshold raising strategies are also applied at different stages of the algorithm to efficiently mine top-k HUIs. 
\par
\begin{figure}[!b]
	\sidecaption[t]
	\includegraphics[height=4cm,width=7.5cm,page=1]{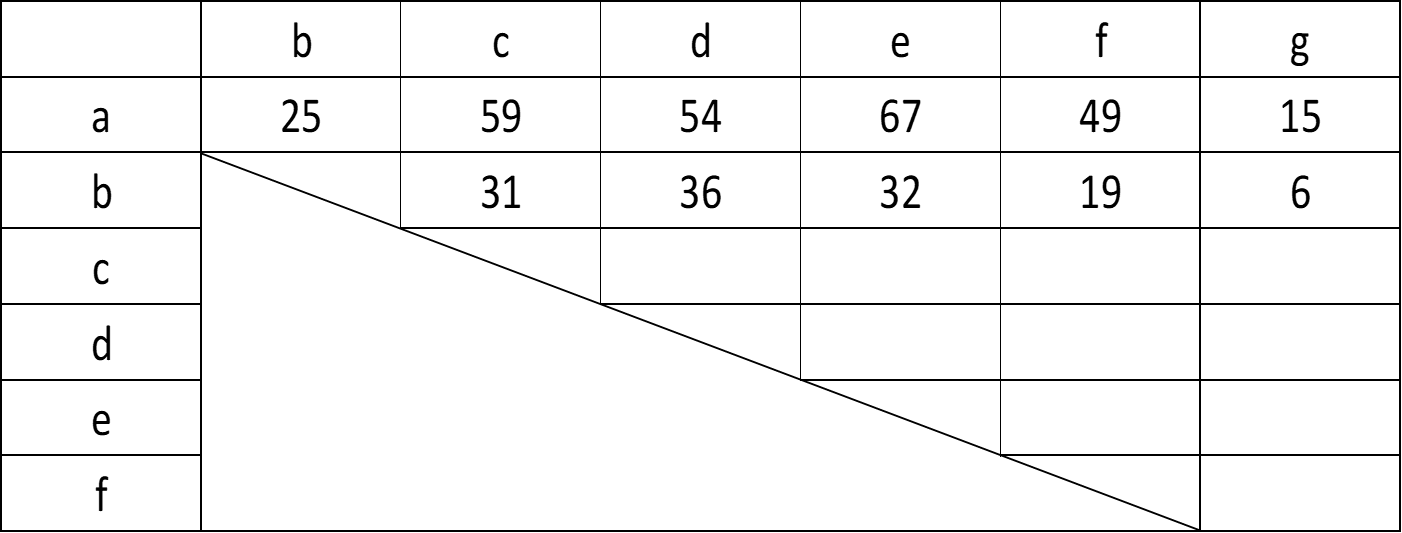}
	\caption{PE matrix for the sample database}
	\label{fig:peMatrix}
\end{figure}

 
The TKU algorithm starts with a $\delta$ value of zero and gradually raises the threshold at different stages of mining. During the first scan of the database, a Pre-evaluation (PE) matrix is constructed to raise the $\delta$ value before the UP-Tree construction. This is done to avoid construction of a full UP-Tree using a $\delta$ value of zero and improve the overall performance of mining. PE matrix contains the lower bounds for the utility of certain 2-itemset pairs. The 2-itemset pairs of a PE matrix are a combination of the first item in any transaction and one of the remaining items in the transaction. For example, when the transaction $T_{1}$ is scanned, the utility of 2-itemset pairs $ac, ad, ae, af$ are accumulated in PE matrix. Similarly, from transaction $T_{2}$, the utility of 2-itemset pairs $ac, ae, ag$ are accumulated. At the end of the first scan of the database, the PE matrix shown in Figure~\ref{fig:peMatrix} is generated.

\begin{strategy} \textit{PE: Threshold Raising Strategy.} \label{strategy:peStrategy}
The PE matrix holds the utility lower bound of certain 2-itemsets. If there are at least K itemsets in the PE matrix, the $\delta$ value can be raised to the Kth highest utility value \cite{wu2012miningTKU}. 
\end{strategy}
The PE threshold raising strategy is used to increase the value of $\delta$ before constructing the UP-Tree. Assuming $K=6$, the $\delta$ value can be raised from 0 to 32 for the running example. 

\begin{figure}[!t]
	\includegraphics[height=8cm,width=11.7cm,page=1]{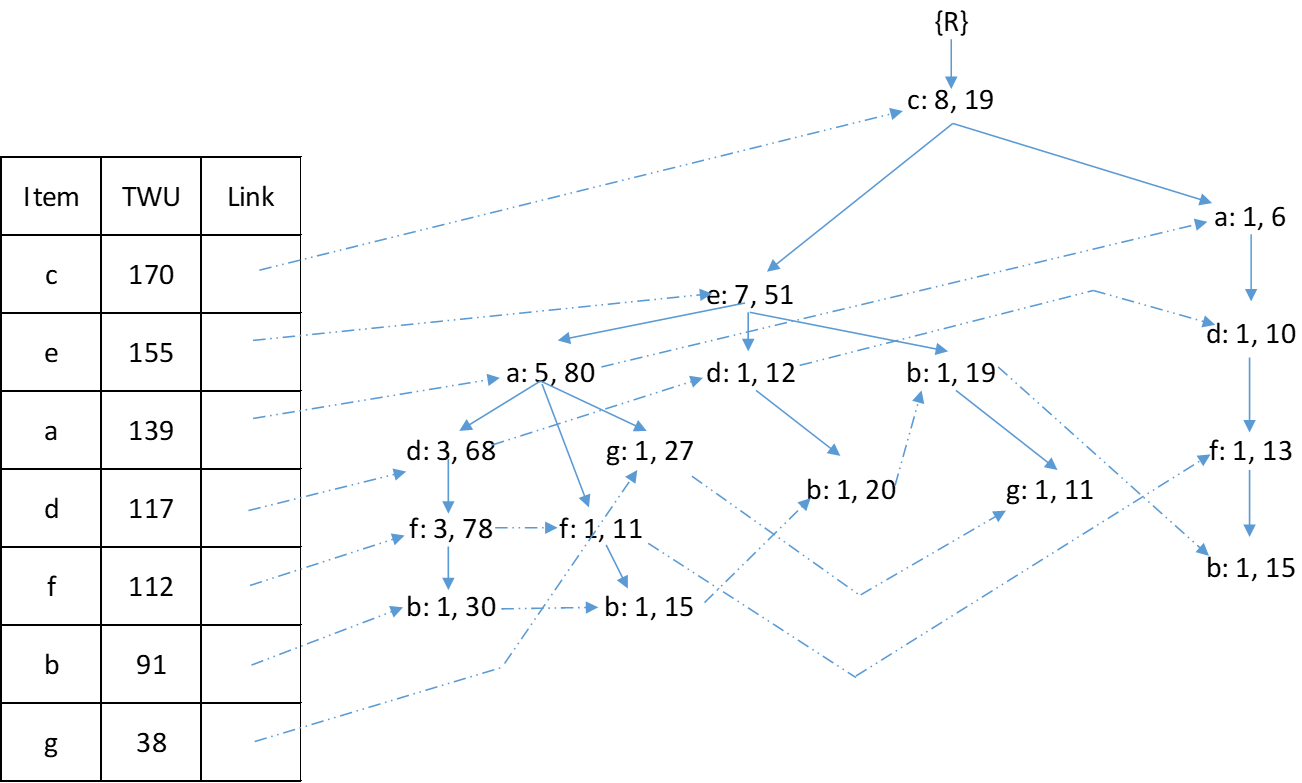}
	\caption{An UP-Tree for the sample database}
	\label{fig:upTree}
\end{figure}

The TKU algorithm then performs a second scan of the database to construct an UP-Tree \cite{Tseng2010UPGrowth}. The unpromising items are filtered using the DGU property (refer to Property~\ref{property:dgu}) during the UP-Tree construction. An UP-Tree consists of two parts: the header table and the actual tree. The header table holds the item information, TWU values and the link to the tree structure. The items in the header table are maintained in a TWU descending order. The tree is constructed by iterating through all the transactions in the database. When a particular transaction is scanned, the items are first sorted in descending order of their TWU values, and then nodes are created (or updated) in the UP-tree. Each node in the tree holds information about the itemset name, support count and the item utilities. For example, the item c has the maximum TWU value of 170 and is the first item in the UP-Tree header table. The item c is created as a child of the root of the tree with support count of 8 and utility value of 19. The constructed UP-Tree for the sample database is shown in Figure~\ref{fig:upTree}. The detailed set of steps involved in the UP-Tree construction can be referred in \cite{Tseng2010UPGrowth,wu2012miningTKU}.

\begin{strategy} \textit{NU: Raising the threshold by Node Utilities.} 
If there are at least K nodes in the UP-Tree and the Kth highest node utility value is greater than $\delta$, then $\delta$ can be raised to the Kth highest node utility value \cite{ryang2015topREPT,wu2012miningTKU}. 
\end{strategy}

For the running example, the UP-Tree has 17 nodes (refer to Figure~\ref{fig:upTree}). Assuming K=6, the 6th highest node utility value in the UP-Tree can be identified as 27. However, the identified node value (of 27) is less than the current $\delta$ value (32). Therefore, the threshold value is not raised by applying the NU strategy for the running example. 
 
\begin{strategy} \textit{MD: Raising the threshold by MIU of Descendants.} 
For every descendant node $N_{b}$ under the root's immediate descendant node ($N_{a}$), the minimum utility of a pair of items, i.e., $miu(X=\{N_{a} N_{b}\})$, is computed. If there are at least K such pair of itemsets and the Kth highest utility value is greater than $\delta$, then $\delta$ can be raised to the Kth highest utility value \cite{wu2012miningTKU}.
\end{strategy}
The MD strategy is applied after constructing the UP-Tree. For the running example, $N_{a} = c$ and the pair of itemsets to be evaluated include $ec, ac, dc, fc, bc, gc$. The estimated lower bound values ($miu$) of every pair of itemsets generated from UP-Tree are provided in Table~\ref{table:miuDescendants}. Assuming K=6, the Kth highest value is 6 which is lower than the current $\delta$ value (32). Hence, the threshold value is not raised by applying the MD strategy for the running example. 

\begin{table}[!h]
	\renewcommand{\arraystretch}{1.2}
	\caption{MIU values of descendants of node c}
	\label{table:miuDescendants}
	%\centering
	\begin{tabular}{>{\centering}p{2.4cm} >{\centering}p{1.2cm} >{\centering}p{1.2cm} >{\centering}p{1.2cm} >{\centering}p{1.2cm} >{\centering}p{1.2cm} >{\centering}p{1.2cm} >{\centering}p{1.2cm}}
		\hline
		Itemset & e & a & d & f & b & g \tabularnewline 
		\hline
		miu & 28 & 36 & 15 & 10 & 15 & 6 \tabularnewline
		\hline
	\end{tabular}
\end{table}

The potential (or the candidate) top-k high utility itemsets (PKHUIs) are mined from the generated UP-Tree. During the mining process, the MC strategy is iteratively applied to quickly raise the threshold $\delta$ value and improve the efficiency of mining. 

\begin{strategy} \textit{MC: Raising the threshold by MIU of Candidates.} 
	If there at least K candidate itemsets and the Kth highest $miu$ of a candidate itemset $X$ is greater than $\delta$, then $\delta$ can be raised to the Kth highest $miu$ value \cite{ryang2015topREPT,wu2012miningTKU}.
\end{strategy}

After the potential top-k HUIs are generated in the first phase, the TKU algorithm identifies all the top-k HUIs. The SE strategy is applied during this phase to raise the $\delta$ value and improve the efficiency of mining top-k HUIs. 

\begin{strategy} \textit{SE: Sorting candidates and Raising the threshold by the exact utility of candidates.} 
	If there are at least K high utility itemsets and the Kth highest utility value of an itemset is greater than $\delta$, then the $\delta$ value can be raised to the Kth highest utility value \cite{ryang2015topREPT,wu2012miningTKU}. 	
\end{strategy}

\par In summary, the TKU algorithm mines all the top-k HUIs in two phases and applies five different threshold raising strategies. The PE strategy is applied during the first scan of phase one. The NU and MD strategies are applied during the second scan of phase one. Finally, the SE strategy is applied during the second phase to efficiently identify all the top-k HUIs. The TKU algorithm also uses four different pruning properties, namely, Decreasing Global Unpromising (DGU) items \cite{Tseng2013UPGrowthPlus,Tseng2010UPGrowth}, Decreasing Global Node (DGN) utilities \cite{Tseng2013UPGrowthPlus,Tseng2010UPGrowth}, Discarding Local Unpromising (DLU) items \cite{Tseng2013UPGrowthPlus,Tseng2010UPGrowth}, and Decreasing Local Node (DLN) utilities \cite{Tseng2013UPGrowthPlus,Tseng2010UPGrowth} at different stage of mining to efficiently mine top-k HUIs. The detailed pseudo-code for the TKU algorithm can be referred in \cite{tseng2016efficientTKO}.

\subsubsection{REPT algorithm} \label{sec:reptAlgorithm}
REPT \cite{ryang2015topREPT} is another two phase method for mining top-k HUIs. The overall functioning of the algorithm is similar to that of the TKU algorithm described earlier. In the first phase, the algorithm constructs an UP-Tree and generates potential top-k HUIs (PKHUIs). Subsequently, in the second phase, the final top-k HUIs are filtered from the generated PKHUIs by computing the exact utilities.  
\par
\begin{figure}[!t]
	\sidecaption[t]
	\includegraphics[height=4cm,width=7.5cm,page=1]{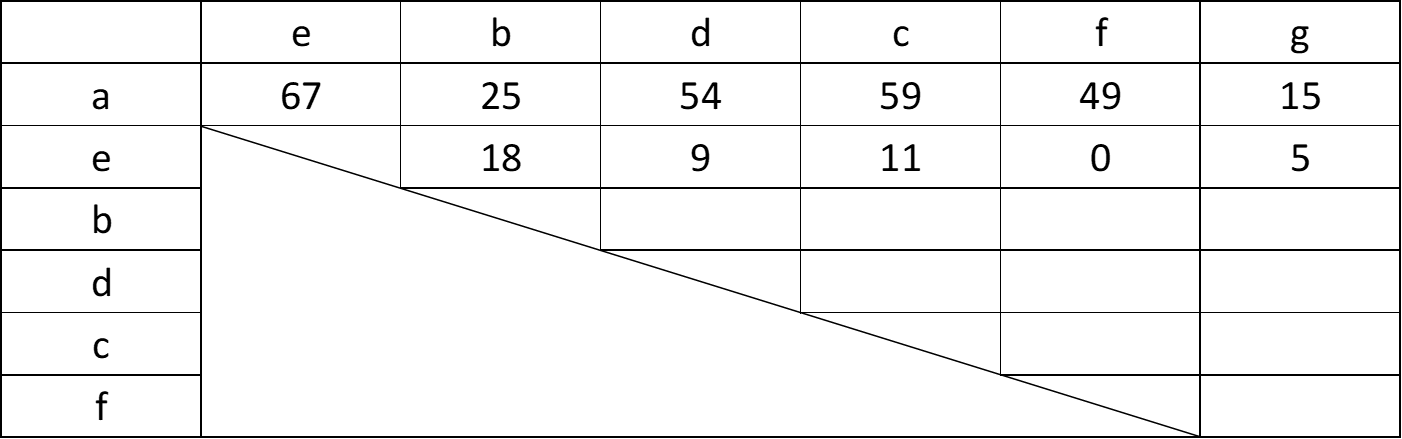}
	\caption{PMUD matrix for the sample database}
	\label{fig:pmudMatrix}
\end{figure}

 
During the first scan of the database, the REPT algorithm constructs a Pre-evaluation matrix with utility descending order (PMUD). The PMUD matrix is similar to the PE matrix used in the TKU algorithm. The key difference lies in the nature of 2-itemsets maintained in the matrix. Unlike TKU algorithm, the first item (of the 2-itemset) is chosen as the item in the transaction with maximum external utility value. For example, the transaction $T_{4}$ has items $b, c, d, e$ and the item with maximum $EU$ value is $e$. Hence, the pair of items $eb, ed, ec$ are generated and stored. On the other hand, the TKU algorithm generates the pair of items $bc, bd, be$ and stores them in the PE matrix. For the running example, the PMUD matrix generated after processing all transactions in the database is provided in Figure~\ref{fig:pmudMatrix}.  

\begin{strategy} \textit{PUD: Raising threshold based on pre-evaluation with utility descending order.} 
	The PMUD matrix holds the utility lower bound value of certain 2-itemsets. If there are at least K itemsets in the PMUD matrix, the $\delta$ value can be raised to the Kth highest utility value \cite{ryang2015topREPT}. 
\end{strategy}

The PUD threshold raising strategy is used to increase the value of $\delta$ before constructing the UP-Tree. Assuming $K=6$, the $\delta$ value can be raised from 0 to 18 for the running example. Though the $PUD$ strategy shows minimal increase in $\delta$ value compared to the $PE$ strategy for the running example, the $PUD$ strategy was found to work well for most of the benchmark datasets \cite{ryang2015topREPT}.  
\par 
After raising the $\delta$ value using the $PUD$ strategy, the REPT algorithm applies another strategy, named RIU, to further increase the value of $\delta$. The strategy involves increasing the value of $\delta$ based on real item utilities. 

\begin{strategy} \textit{RIU: Raising threshold based on real item utilities.} \label{strategy:riu}
	If there are at least K items and the Kth highest RIU value (i.e., $RIU_{k}$) is greater than $\delta$, then $\delta$ value can be raised to $RIU_{k}$.
\end{strategy}
For the sample database with K=6, $RIU_{6} = 14$ which is lower than the current $\delta$ value of 18 (obtained after applying the $PUD$ strategy). Hence, the threshold value is not raised by applying the RIU strategy for the running example. 

\par The REPT algorithm also proposed a new RSD matrix to maintain exact utility information for certain 2-itemsets. The items for the matrix are chosen based on the support value of individual items computed in the first scan of the database. The algorithm first sorts the items in descending order of 1-item support values. Then, it selects N/2 items with highest support values and N/2 items with lowest support values, where N is the desired number of promising items to be explored. 
\par
\begin{figure}[!t]
	\sidecaption[t]
	\includegraphics[height=2.4cm,width=6cm,page=1]{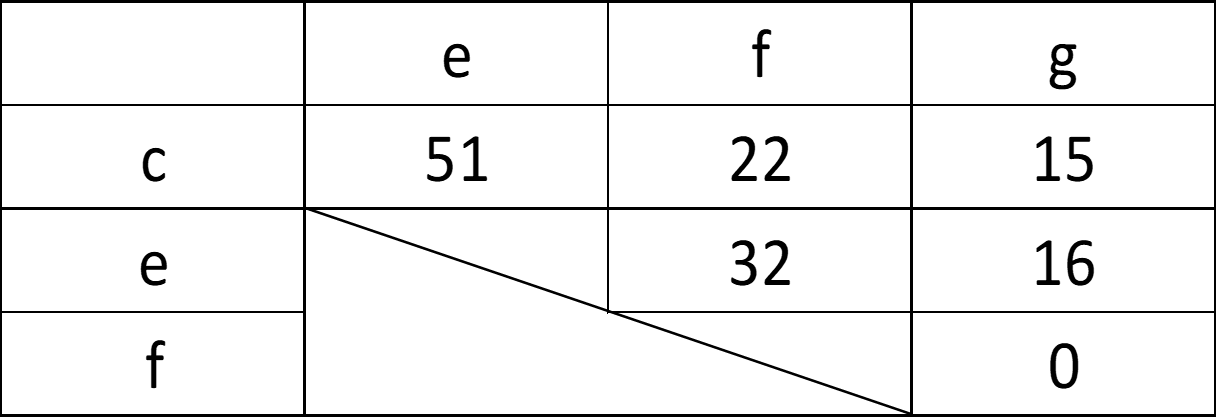}
	\caption{RSD matrix for the sample database}
	\label{fig:rsdMatrix}
\end{figure}

 
For the running example, the item reordering as per the support descending order is: $\{c, e, a, b, d, f, g\}$. Assuming N=4, the top two items with highest support are c and e. The two items with lowest support are f and g. These four items (c, e, f and g) are used to construct the RSD matrix. The entries in the RSD matrix are initially set to zero and then updated during the second scan of the database. While each and every transaction is processed, the utility values of pair of selected items are updated in the RSD matrix. For example, when transaction $T_{1}$ is processed, the 2-itemset pairs $ce, cf, ef$ are generated. Similarly, when transaction $T_{7}$ is processed, the 2-itemset pair $cf$ is generated, and its utility value is updated in the matrix. The complete RSD matrix for the sample database is shown in Figure~\ref{fig:rsdMatrix}. 

\begin{strategy} \textit{RSD: Raising threshold with items in support descending order.} \label{strategy:rsd}
	If there are at least K itemsets in the RSD matrix and the Kth highest value is greater than the current $\delta$ value, then the $\delta$ value can be increased to the Kth highest value in the RSD matrix \cite{ryang2015topREPT}. It is to be noted that the size of the RSD matrix is dependent on a user specified parameter $N$, where $N$ indicates the desired number of promising items to be explored. 
\end{strategy}

With K=6, the Kth highest value in RSD matrix is zero. Since the Kth highest value is lower than the current threshold value (18), no change in the value of $\delta$ is made. It is to be noted that the RSD strategy is similar to the MD strategy proposed in TKU algorithm. While the MD strategy uses the minimum estimated utility value of 2-itemsets, the RSD strategy uses the exact utility values of 2-itemsets. It is quite intuitive to understand that the RSD strategy is likely to be better (compared to MD \cite{tseng2016efficientTKO}) as it uses exact utility values. The experimental evaluation of the RSD strategy reported in \cite{ryang2015topREPT} clearly demonstrates its benefits.

\begin{strategy} \textit{SEP: Raising threshold by sorting candidates and computing exact utilities} 
	This strategy is an extension of the SE strategy introduced in TKU algorithm. The SEP strategy primarily uses the real item utilities and the RSD matrix to compute exact item utilites (instead of estimated utilites) and improve the efficiency of mining. 
\end{strategy}

\par 
In summary, the REPT algorithm mines all the top-k HUIs in two phases and applies six different threshold raising strategies. In phase one, the algorithm applies the PUD and RIU strategies before constructing the UP-Tree. The UP-Tree is then constructed and the potential top-k HUIs (PKHUIs) are mined. The NU and RSD strategies are then applied to raise the threshold values ($\delta$) and improve the efficiency of mining PKHUIs. Finally, in phase two, the top-k HUIs are determined from the set of PKHUIs. The SEP strategy is applied in this phase to efficiently identify all the top-k HUIs. REPT algorithm also uses five different pruning properties, namely, TWDC, Decreasing Global Unpromising (DGU) items \cite{Tseng2013UPGrowthPlus,Tseng2010UPGrowth}, Decreasing Global Node (DGN) utilities \cite{Tseng2013UPGrowthPlus,Tseng2010UPGrowth}, Discarding Local Unpromising (DLU) items \cite{Tseng2013UPGrowthPlus,Tseng2010UPGrowth}, and Decreasing Local Node (DLN) utilities \cite{Tseng2013UPGrowthPlus,Tseng2010UPGrowth} at different stage of mining to efficiently mine top-k HUIs. The detailed pseudo-code for the REPT algorithm can be referred in \cite{ryang2015topREPT}.

\par 
From the foregoing discussions on two-phase methods, it is evident that REPT \cite{ryang2015topREPT} is the state-of-the-art two-phase top-k HUI mining method. The authors present several new threshold raising strategies and demonstrate the superiority of their method over the TKU method \cite{wu2012miningTKU}. One of the drawbacks of REPT, however, is the need for specification of additional parameter $N$ to effectively use the RSD strategy (refer to strategy~\ref{strategy:rsd}). Tseng et al \cite{tseng2016efficientTKO} study the impact of varying $N$ value on the performance of REPT algorithm. Their experiments reveal that the proper choice of $N$ is important for effective use of REPT method for mining top-k HUIs. The authors \cite{duong2016efficientKHMC,tseng2016efficientTKO} argue that this will be quite challenging especially for users who lack domain expertise and might lead to lot of trial and error in tuning the algorithm performance.

\subsection{One-phase methods}
\label{subsec:m2}

The two-phase methods often generate too many candidate top-k HUIs before mining the actual top-k HUIs. The candidate generation process is quite expensive, especially on dense and long transactional databases. More recent methods in the literature address the limitations of two-phase methods by completely avoiding the expensive candidate generation process. These methods work in single phase and generate all the top-k HUIs. The two prominent methods that work in one-phase include TKO \cite{tseng2016efficientTKO} and KHMC \cite{duong2016efficientKHMC}. Both of these methods rely on a vertical database representation structure, named utility list \cite{Liu2012HUIMiner}, and use a tree enumeration method to effectively mine top-k HUIs in a single phase. 
\par In this section, we first introduce a few key definitions commonly used in the context of one-phase top-k HUIs. Subsequently, we discuss the one-phase top-k HUI mining methods (TKO and KHMC) in detail. 

\begin{definition}\label{def:ordHeuristic}
	\textit{(Ordering of items).} The items in the transaction database are processed using total order $\prec$  such that the items are sorted in TWU ascending order. This ordering heuristic is commonly used in one-phase HUI mining methods in the literature. 
\end{definition}
\par For the running example, the ordering of items are: g $\prec$ b $\prec$ f $\prec$ d $\prec$ a $\prec$ e $\prec$ c. The individual transactions in the database are also ordered as per this heuristic and the ordering for the sample database is shown in Table~\ref{table:ProcessedDB}.   

\begin{table}[!h]
	\renewcommand{\arraystretch}{1.2}
	\caption{Ordered transaction database}
	\label{table:ProcessedDB}
	%\centering
	%\begin{tabular}{cllc}
	\begin{tabular}{>{\centering}p{1.2cm} p{3cm} p{3cm} >{\centering}p{2cm}}
		\hline
		TID & Transaction & Utility (U) & TU \tabularnewline
		\hline
		$T_{1}$ & f, d, a, e, c & 2, 2, 5, 6, 1 & 16 \tabularnewline
		$T_{2}$ & g, a, e, c & 5, 10, 6, 6 & 27 \tabularnewline 
		$T_{3}$ & b, f, d, a, e, c & 4, 5, 12, 5, 3, 1 & 30 \tabularnewline
		$T_{4}$ & b, d, e, c & 8, 6, 3, 3 & 20 \tabularnewline
		$T_{5}$ & g, b, e, c & 2, 4, 3, 2 & 11 \tabularnewline 
		$T_{6}$ & f, d, a, e, c & 3, 6, 15, 9, 3 & 36 \tabularnewline 
		$T_{7}$ & b, f, d, a, c & 2, 3, 4, 5, 1 & 15 \tabularnewline 
		$T_{8}$ & b, f, a, e, c & 4, 1, 5, 3, 2 & 15 \tabularnewline 
		\hline
	\end{tabular}
\end{table}

\begin{figure}[!h]
	\includegraphics[height=6cm,width=11.7cm,page=1]{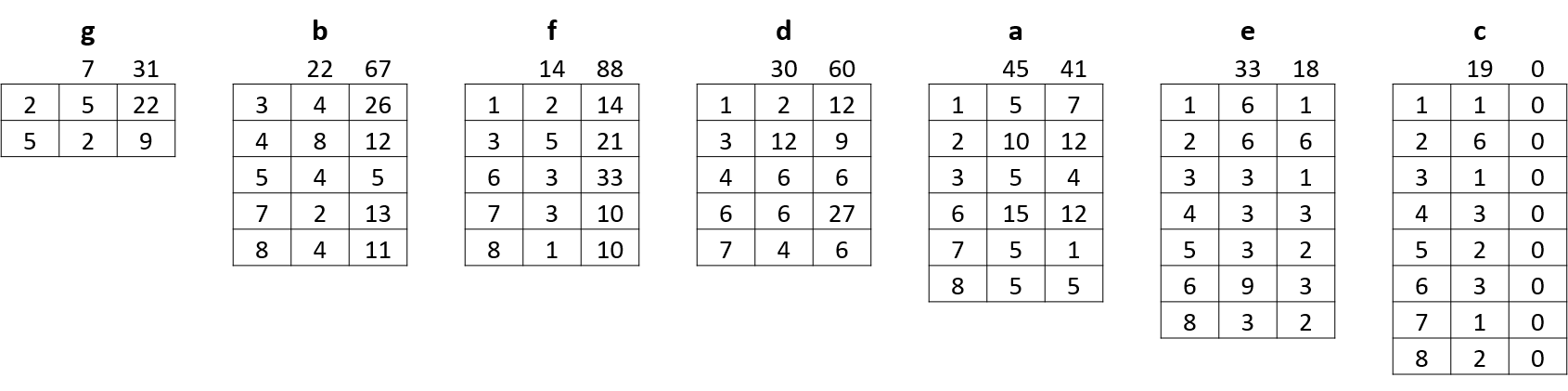}
	\caption{Utility list for the sample database}
	\label{fig:utilityList}
\end{figure}

\begin{definition}\label{def:gX}
	The tidset of an itemset $X$, denoted as $g(X)$, is defined as 
	\begin{equation}
		g(X) = \{tid | X \subseteq T_{tid} \hspace*{.5em} and \hspace*{.5em} T_{tid} \in D\}
	\end{equation}
\end{definition}
\par For example, $g(f) = \{1, 3, 6, 7, 8\}$ since the itemset $f$ is present in transactions 1, 3, 6, 7 and 8 (refer to Table~\ref{table:SampleDB}). 

\begin{definition}\label{def:ulist}
The utility list of an itemset $X$, denoted as $UL(X)$, is a data structure that holds: (1) summary information about utility and remaining utility of items i.e., $U(X)$ and $RU(X)$, and (2) transactional level information (element) in the form of triplets $<tid, U(X, T_{j}), RU(X, T_{j}))>$. The transaction information is maintained for all transactions $T_{j} \in g(X)$. 
\end{definition}

\par 
Figure~\ref{fig:utilityList} shows the utility list for the ordered transaction database in Table~\ref{table:ProcessedDB}. For example, the itemset $g$ contains the summary information as 7 and 31 since $U(g) = U(g, T_{2}) + U(g, T_{5}) = 5 + 2 = 7$ and $RU(g) = RU(g, T_{2}) + RU(g, T_{5}) = 22 + 9 = 31$. The individual transaction level information are maintained as separate entries in the utility list. Similarly, itemset $d$ occurs in transactions 1, 3, 4, 6 and 7. Hence, there are 5 entries in the utility list for itemset $d$. The summary information for itemset $d$ contains the value as 30 and 60 since $U(d) = 30$ and $RU(d) = 60$.

\begin{definition} \textit{(Z-element)} 
	An element (or transaction entry) in the utility list is called an Z-element iff its remaining utility value is equal to zero. Otherwise, the element is called an NZ-element. The set of all NZ-elements in the utility list of X is denoted as NZE(X).
\end{definition}

\par For the running example, the NZE(gb) = $\{<5,6,5>\}$, NZE(g) = $\{<2,5,22>, <5,2,9>\}$ and NZE(c) = $\{\}$.

\begin{definition} \textit{(Non-zero element utilities)} 
	The non-zero element utilities of an itemset $X$, denoted as NZEU(X), is defined as the sum of the utilities of non-zero elements in the utility list of $X$. 
\end{definition}

\par For the running example, NZEU(gb) = 6, NZEU(g) = 7 and NZEU(c) = 0.

\subsubsection{TKO algorithm}
Wu et al \cite{wu2012miningTKU} extend their TKU algorithm and introduce a new one-phase method named TKO in \cite{tseng2016efficientTKO}. The TKO algorithm uses a utility list data structure \cite{Liu2012HUIMiner} for maintaining itemset information during the mining process. 
\par 
The algorithm starts with a $\delta$ value of zero and initally scans the database to compute the TWU and utility values of items. During the first scan, a pre-evaluation (PE) matrix is also constructed to raise the $\delta$ value. The PE matrix construction process is similar to the one described earlier in section~\ref{algo:tkuAlgorithm}. The PE matrix for the running example is also provided in Figure~\ref{fig:peMatrix}. After the completion of the first scan, the $\delta$ value is raised by applying the PE threshold raising strategy (refer to strategy~\ref{strategy:peStrategy}).

\par 
TKO algorithm then scans the database again and constructs the 1-item utility lists. During the scan, the DGU property (refer to property~\ref{property:dgu}) is applied to filter unpromising items from further processing. The items in each transaction are also sorted as per the ordering heuristic (refer to definition~\ref{def:ordHeuristic}). For the sample database in Table~\ref{table:SampleDB}, the ordered set of items in individual transactions are provided in Table~\ref{table:ProcessedDB}. As the individual transactions are scanned from the database, the utility lists are iteratively constructed. The complete 1-item utility list for the running example is shown in Figure~\ref{fig:utilityList}. 

\begin{algorithm}[!t]
	\renewcommand{\thealgorithm}{}%%for disabling numbering
	\caption{\textbf{1} TKO Algorithm: Search-Tree-Exploration}
	\label{alg:explore-search-tree}
	{\bfseries Input:} $R$, the $UL$ of itemset R, \\ 
	\hspace*{2.2em} $ULs$, the set of $UL$s of all R's 1-extensions, \\ 
	\hspace*{2.2em} $\delta$, the current minimum utility threshold value \\
	\hspace*{2.2em} TopK-CI-List, a list for storing candidate itemsets \\
	{\bfseries Output:} all Top-K HUIs with prefix $R$ 
	\begin{algorithmic}[1]
		\FOR[\textit{//Explore Search Tree}]{each utility list $X$ in $ULs$}
		\STATE{\textbf{if} $U(X) \geq \delta$ \textbf{then} $\delta = RUC(X, $TopK-CI-List$)$}
		\IF[\textit{//U-Prune \cite{krishnamoorthy2015pruning,Liu2012HUIMiner}}]{$U(X)+ RU(X) \geq \delta$} 
		\STATE{$exULs$ $\leftarrow$ \{\}}
		\FOR{each utility list $Y$ after $X$ in $ULs$}
		\STATE{$UL(XY)$ = ConstructUL($R, X, Y$) \COMMENT{\textit{//refer to \cite{Liu2012HUIMiner,tseng2016efficientTKO} for details}}}
		\STATE{$exULs$ = \{$exULs \hspace{0.8em} \cup \hspace{0.8em} UL(XY)$\}}
		\ENDFOR
		\STATE{Explore-Search-Tree($X$,$exULs$,$\delta$, TopK-CI-List)}   
		\ENDIF
		\ENDFOR   
	\end{algorithmic}
\end{algorithm}

\par The generated 1-item utility lists are used to explore the search space and mine the top-k HUIs. A min heap structure named TopK-CI-List is maintained to store the current set of top-k HUIs during the search process. The recursive search exploration process is exactly similar to the standard list based approach like HUI-Miner \cite{Liu2012HUIMiner}. The pseudo-code for the search exploration process in TKO algorithm is provided in Algorithm 1. The key differences in the search exploration process (compared to \cite{Liu2012HUIMiner}) are on three aspects: (1) RUC threshold raising strategy, (2) RUZ pruning property, and (3) EPB property. Each of these aspects and their utility in top-k HUI mining are described next. 

\begin{strategy} \textit{RUC: Raising the threshold by the Utilities of Candidates.} \label{strategy:ruc}
	If there are at least K high utility itemsets in TopK-CI-List structure and the Kth highest utility value of an itemset is greater than $\delta$, then the $\delta$ value can be raised to the Kth highest utility value \cite{tseng2016efficientTKO}. 	
\end{strategy}

\par RUC strategy is similar, in principle, to the SE and SEP strategies used in \cite{ryang2015topREPT,wu2012miningTKU}. The candidate top-k HUIs are maintained in a priority queue structure, named TopK-CI-List. The entries in the queue are updated when a new candidate with higher utility value is observed. The RUC strategy helps in raising the threshold value and improving the performance of top-k HUI mining. This strategy is incorporated as part of the RUC function (refer to step 2 of Algorithm 1). The function updates the TopK-CI-List and revises the $\delta$ value by applying the RUC strategy.

\begin{propertyNew} \textit{RUZ: Reducing estimated utility values by using Z-elements.} \label{property:ruz} \\ If $NZEU(X) + RU(X) < \delta$, then all extensions of $X$ are not top-k HUIs \cite{tseng2016efficientTKO}. 
\end{propertyNew}

\par This property is a simple extension to the U-Prune property \cite{krishnamoorthy2015pruning,Liu2012HUIMiner} where the zero elements are excluded from the total utility computations. The TKO algorithm employs the RUZ pruning property to improve the performance of mining. More specifically, the line 3 of Algorithm 1 will be replaced with this RUZ property to prune non-promising candidates during the search tree exploration process.

\begin{propertyNew} \textit{EPB: Exploring the most Promising Branches first.} \label{property:epb} 
	The EPB strategy primarily processes the most promising candidates with highest utility values first. More specifically, the utility list extensions ($ULs$ in Algorithm 1) of a given prefix ($R$ in Algorithm 1) are explored in decreasing order of their estimated utility value. The estimated utility value is determined as the sum of the utility and remaining utility value of an itemset.
\end{propertyNew}

\par The EPB strategy allows pruning unpromising candidates by quickly raising the threshold value during the mining process. 

\par 
Overall, the TKO algorithm mines all the top-k HUIs in a single phase. It applies two key threshold raising strategies (PE and RUC) and four different pruning properties (DGU, RUZ, EPB, U-Prune \cite{krishnamoorthy2015pruning,Liu2012HUIMiner}) at different stages of the mining process. The authors demonstrate that their one-phase method is superior compared to the baseline two-phase TKU \cite{wu2012miningTKU} and REPT \cite{ryang2015topREPT} methods.

\subsubsection{KHMC algorithm}

\par 
KHMC \cite{duong2016efficientKHMC} is the most recent top-k HUI mining method that adopts a one-phase utility list based approach. The algorithm first scans the database to compute the TWU and utility values of items. The algorithm then applies the RIU strategy \cite{ryang2015topREPT} (refer to strategy~\ref{strategy:riu} in section~\ref{sec:reptAlgorithm}) to increase the $\delta$ value. A second scan of the database is then made to construct the EUCST, CUDM and utility list data structures. The utility list constructed is the same as the one described earlier in TKO algorithm. The utility list for the running example is provided in Figure~\ref{fig:utilityList}. The details of the EUCST and CUDM are described next.

\begin{definition} \label{def:eucs}
	Estimated Utility Co-occurrence Structure (EUCST) \cite{duong2016efficientKHMC} is a hash map data structure that stores the TWU information of a pair of items. A 2-itemset ($X = \{x_{i}, x_{j}\}$) entry in EUCST is defined as 
	\begin{equation}
	%\begin{gather}
	%\nonumber 
	EUCST(X=\{x_{i}, x_{j}\}) = TWU(X = \{x_{i}, x_{j}\})
	%\end{gather}
	\end{equation} 
\end{definition}

\par 
The EUCST structure proposed in \cite{duong2016efficientKHMC} is an enhancement to the EUCS  structure introduced in \cite{fournier2014fhmEUCP}. The key difference between these two structures are in terms of their underlying implementations. While the EUCS uses a triangular matrix, the EUCST uses a hash map data structure. The latter structure is optimized in terms of space compared to the former EUCS structure. For the running example, $EUCST$ of an itemset $X = \{da\}$ is computed as $EUCST(X = \{da\}) = TWU(\{da\}) = TU(T_{1}) + TU(T_{3}) + TU(T_{6}) + TU(T_{7}) = 16 + 30 + 36 + 15 = 97$.

\begin{definition} \label{def:cudm}
	Co-occurrence Utility Descending order utility Matrix (CUDM) \cite{duong2016efficientKHMC} is a hash map data structure that stores the utility information of a pair of items. A 2-itemset ($X = \{x_{i}, x_{j}\}$) entry in CUDM is defined as 
	\begin{equation}
	CUDM(X=\{x_{i}, x_{j}\}) = U(X = \{x_{i}, x_{j}\})
	\end{equation} 
\end{definition}

\par 
For the running example, $CUDM$ of an itemset $X = \{da\}$ is computed as $CUDM(X = \{da\}) = U(\{ad\}) = U(T_{1}) + U(T_{3}) + U(T_{6}) + U(T_{7}) = 7 + 17 + 21 + 9 = 54$.

\begin{strategy} \textit{CUD: Co-occurrence Utility Descending order threshold raising strategy.} 
	If there are at least K itemsets in CUDM matrix and the Kth highest utility value of an itemset is greater than $\delta$, then the $\delta$ value can be raised to the Kth highest utility value \cite{duong2016efficientKHMC} in CUDM. 	
\end{strategy}

\par 
The CUD strategy was introduced in \cite{duong2016efficientKHMC}. It is used to raise the threshold value ($\delta$) at the end of the second database scan. 

\par 
After raising the threshold using $CUD$ strategy, another new coverage based strategy is applied in KHMC algorithm. The coverage based strategy (COV) is aimed at further increasing the $\delta$ value and improve the performance of mining during subsequent growth stage. 

\begin{definition} (Coverage of an item)
	Let $x$ and $y$ be two single items. The item $y$ is said to cover item $x$ if $g(x) \subseteq g(y)$. The coverage of an item $x$, denoted as $C(x)$, is defined as $C(x) = \{y | y \in I, g(x) \subseteq g(y)\}$.
\end{definition}

\par For the running example, $g(g) = \{2, 5\}, g(a) = \{1,2,3,6,7,8\}, g(b) = \{3,4,5,7, \\8\}, g(e) = \{1,2,3,4,5,6,8\}$ and $g(c) = \{1,2,3,4,5,,7,8\}$. Therefore, $C(g) = \{e, c\}$.

\par The coverage of single items are used to estimate utilities of superset items in \cite{duong2016efficientKHMC}. The estimated utility values of superset items are then stored in a data structure named, COV. The information stored in COV data structure is used to raise the threshold value by applying the COV strategy.

\begin{table*}[!hb]
	\renewcommand{\arraystretch}{1.2}
	\caption{Summary of threshold raising strategies used by top-k HUI mining methods}
	\label{table:thresholdSummary}
	%\centering
	\begin{tabular}{>{\centering}p{1cm} >{\centering}p{2cm} >{\centering}p{2cm} >{\centering}p{2cm} >{\centering}p{2cm} >{\centering}p{2cm}}
		\hline
		\# & Strategy & TKU & REPT & TKO & KHMC \tabularnewline 
		\hline 
		1 & PE & Phase1 & & Phase1 & \tabularnewline 
		2 & NU & Phase1 & Phase1 & &  \tabularnewline 
		3 & MD & Phase1 & & & \tabularnewline 
		4 & MC & Phase1 & Phase1 & & \tabularnewline 
		5 & SE & Phase2 & & & \tabularnewline 
		6 & PUD & & Phase1 & & \tabularnewline 
		7 & RIU & & Phase1 & & Phase1 \tabularnewline 
		8 & RSD & & Phase1 & & \tabularnewline 
		9 & SEP & & Phase2 & & \tabularnewline 
		10 & RUC & & & Phase1 & Phase1 \tabularnewline 
		11 & CUD & & & & Phase1 \tabularnewline 
		12 & COV & & & & Phase1 \tabularnewline 
		\hline
	\end{tabular}
\end{table*}

\begin{strategy} \textit{COV: Coverage threshold raising strategy.} 
	If there are at least K itemsets in COV data structure and the Kth highest utility value of an itemset is greater than $\delta$, then the $\delta$ value can be raised to the Kth highest utility value \cite{duong2016efficientKHMC} in COV. 
\end{strategy}

\par 
The KHMC algorithm applies the CUD and COV strategies at the end of second scan of the database to increase the threshold value ($\delta$). The utility list of single items are also constructed at the end of second scan of the database. The generated 1-item utility lists are used to explore the search space and mine the top-k HUIs. During the search tree exploration process, the RUC strategy (strategy~\ref{strategy:ruc}) is applied to increase the $\delta$ value. Three pruning properties were also applied during the growth stage of mining to improve the overall performance of top-k HUI mining. The utility prune (U-Prune), early abandonment (EA) and transitive extension pruning (TEP) are the three pruning properties used in KHMC algorithm. The U-Prune, EA and TEP properties were primarily inspired from U-Prune \cite{krishnamoorthy2015pruning,Liu2012HUIMiner}, LA \cite{krishnamoorthy2015pruning} and Sub-tree Utility (SU) \cite{zida2015efim} properties used in the past literature. 
\par 
In summary, the KHMC algorithm uses a utility list based approach for mining top-k HUIs in a single phase. It uses four threshold raising strategies (RIU, CUD, COV and RUC) and five pruning properties (TWDC, EUCS, U-Prune, EA and TEP) to effectively mine the top-k HUIs. The authors demonstrate the effectiveness of their method through rigorous experimental evaluation. A summary of threshold raising strategy and pruning properties used by different top-k HUI mining methods are provided in Tables~\ref{table:thresholdSummary} and~\ref{table:pruneSummary}.

\begin{table*}[!hb]
	\renewcommand{\arraystretch}{1.2}
	\caption{Summary of pruning properties used by top-k HUI mining methods}
	\label{table:pruneSummary}
	%\centering
	\begin{tabular}{>{\centering}p{2cm} >{\centering}p{2cm} >{\centering}p{2cm} >{\centering}p{2cm} >{\centering}p{2cm}}
		\hline
		Property & TKU & REPT & TKO & KHMC \tabularnewline 
		\hline 
		TWDC & & Y &  & Y \tabularnewline 
		DGU & Y & Y & Y &  \tabularnewline 
		DGN & Y & Y & & \tabularnewline 
		DLU & Y & Y &  & \tabularnewline 
		DLN & Y & Y & & \tabularnewline 
		RUZ & & & Y & \tabularnewline 
		EPB & & & Y & \tabularnewline 
		U-Prune & & & Y & Y \tabularnewline 
		EUCS & & & & Y \tabularnewline 
		EA & & & & Y \tabularnewline 
		TEP & & & & Y \tabularnewline 
		\hline
	\end{tabular}
\end{table*}

\section{Performance analysis of state-of-the-art Top-K HUI mining methods}
\label{sec:4}

In this section, we experimentally evaluate the performance of the state-of-the-art top-k HUI mining methods. As the one-phase methods are proven to be superior compared to the two-phase methods, we primarily analyze the performance of one-phase methods. More specifically, we analyze the performance of TKO \cite{tseng2016efficientTKO} and KHMC \cite{duong2016efficientKHMC} methods. 

\subsection{Experimental design}
\label{subsec:e1}

We implemented the two algorithms (TKO and KHMC) by extending the SPMF open source data mining library \cite{Fournier2014SPMF}. All our experiments were performed on a Dell workstation having Intel Xeon 3.7GHz processor with 64GB of main memory, 8GB java heap size and running a linux operating system. We evaluated the performance of the algorithms on four sparse (chain, kosarak, retail, accidents) and four dense (pumsb, mushroom, connect, chess) benchmark datasets. The details of datasets used in our experiments are shown in Table~\ref{table:dataset}. All the datasets, except chain, were downloaded from \cite{Fournier2014SPMF}. The chain dataset was downloaded from \cite{Pisharath2005NUMineBench}. 
\par 
We implemented two versions of KHMC algorithm using the base version shared to us by the authors \cite{duong2016efficientKHMC}. One version works with the TEP property \cite{duong2016efficientKHMC} enabled and another without it. Two different versions were required since the TEP requires a pure depth first implementation (in contrast to standard utility list implementation \cite{Liu2012HUIMiner}) for itemset tree exploration. We observed that the KHMC algorithm without TEP offers the best performance results. Hence, we used this implementation in all our experimental evaluations. 

\begin{table}[!h]
	\renewcommand{\arraystretch}{1.2}
	\caption{Dataset characteristics}
	\label{table:dataset}
	\centering 
	%\begin{tabular}{lrrrr}
	\begin{tabular}{P{2cm}X{2cm}X{2.4cm}X{2cm}X{2.8cm}}
		\hline\noalign{\smallskip}
		\multicolumn{1}{c}{Dataset} & \multicolumn{1}{c}{\#Trans} & \multicolumn{1}{c}{\#Items (I)} & \multicolumn{1}{c}{AvgLen(L)} & \multicolumn{1}{c}{Density (L/I) \%} \\
		\hline 
		chain & 1112949 & 46086 & 7.3 & 0.0158 \\ 
		kosarak & 990002 & 41270 & 8.1 & 0.0196 \\ 
		retail & 88162 & 16470 & 10.3 & 0.0625 \\ 
		pumsb & 49046 & 2113 & 74 & 3.5021 \\
		accidents & 340183 & 468 & 33.8 & 7.2222 \\
		mushroom & 8124 & 119 & 23 & 19.3277 \\
		connect & 67557 & 129 & 43 & 33.3333 \\
		chess & 3196 & 75 & 37 & 49.3333 \\
		\noalign{\smallskip}\hline\noalign{\smallskip}
	\end{tabular} 
\end{table}

\subsection{Experimental results}
\label{subsec:e2}

\begin{figure}[!t]
	\includegraphics[height=8cm,width=11cm,page=1]{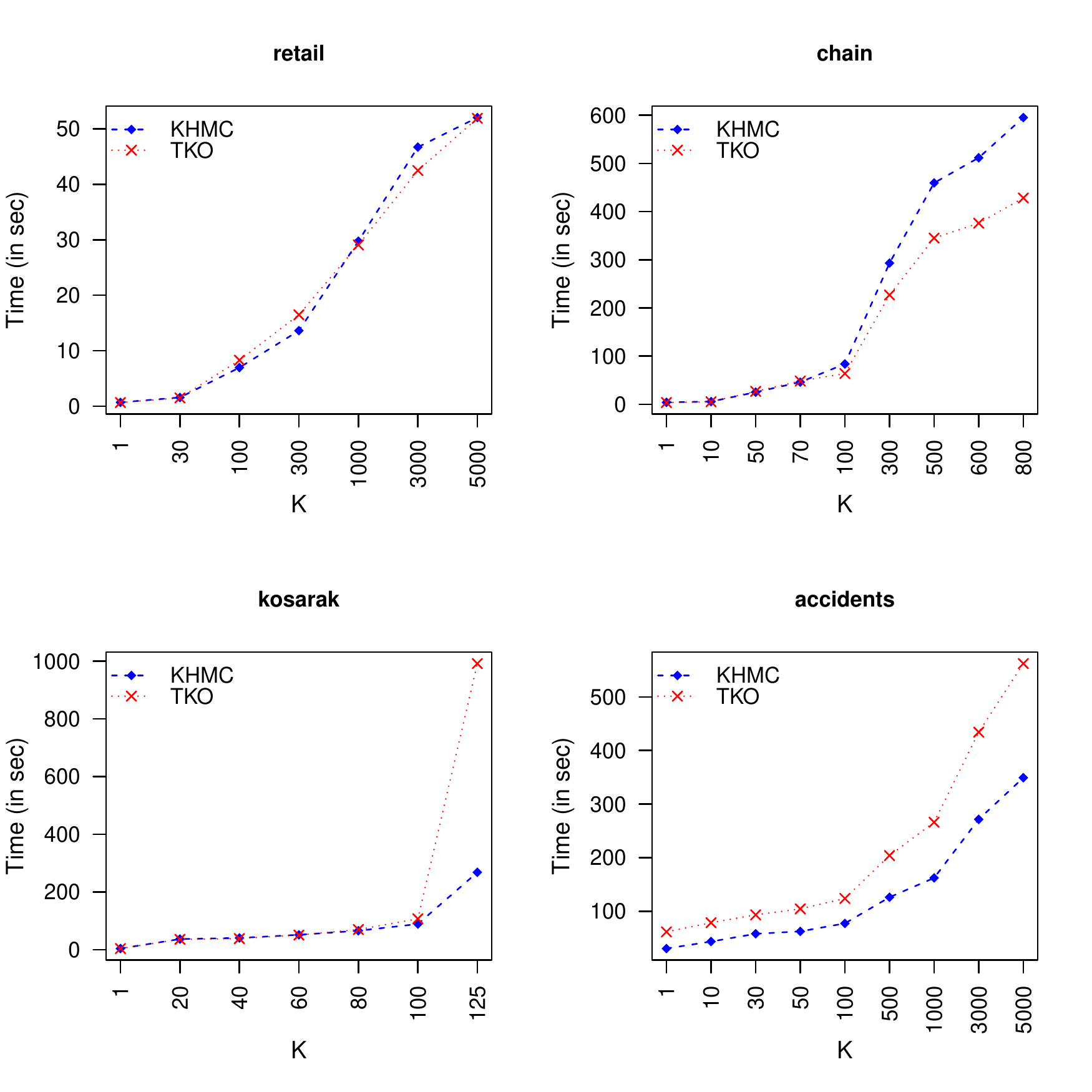}
	\caption{Runtime analysis of KHMC and TKO on sparse datasets}
	\label{fig:runtimeSparse}
\end{figure}

\begin{figure}[!h]
	\includegraphics[height=7.5cm,width=11cm,page=2]{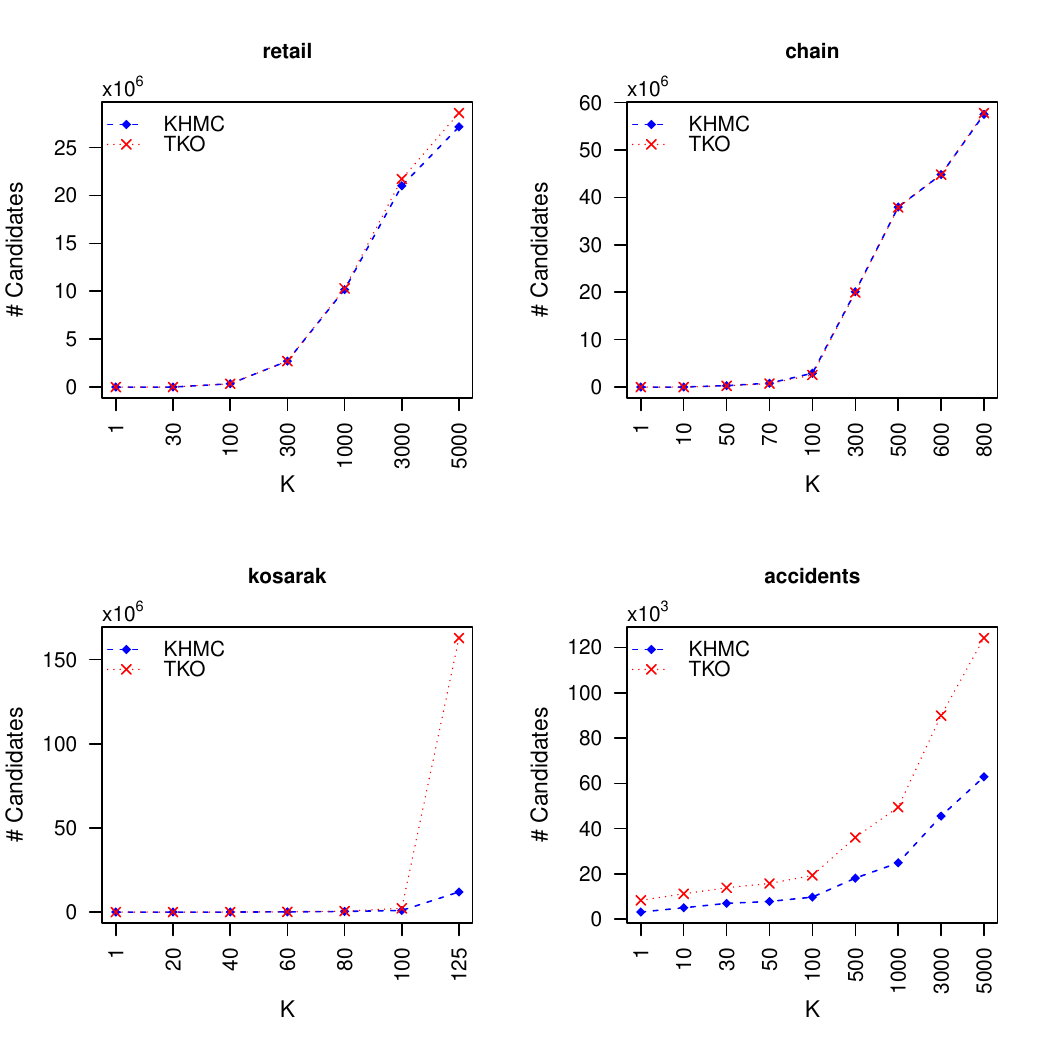}
	\caption{Number of candidates generated on sparse datasets}
	\label{fig:candidatesSparse}
\end{figure}

We analyze the performance of TKO and KHMC on sparse as well as dense datasets. In the first set of experiments, we study the performance of algorithms on sparse datasets. Figure~\ref{fig:runtimeSparse} provides the results of our runtime experiments at varying levels of $K$. The results reveal that KHMC works better on kosarak and accidents dataset. As the value of $K$ is increased, the performance of TKO algorithm degrades significantly. On the more sparser retail and chain dataset, the TKO algorithm was found to perform better, though the margin of difference is quite small. Moreover, the total number of candidates generated by these algorithms on retail and chain dataset was observed to be very similar. This is evident from the experimental analysis results shown in Figure~\ref{fig:candidatesSparse}. These results indicate that the KHMC algorithm works well on most of the sparse benchmark datasets studied.

\begin{figure}[!t]
	\includegraphics[height=8cm,width=11cm,page=3]{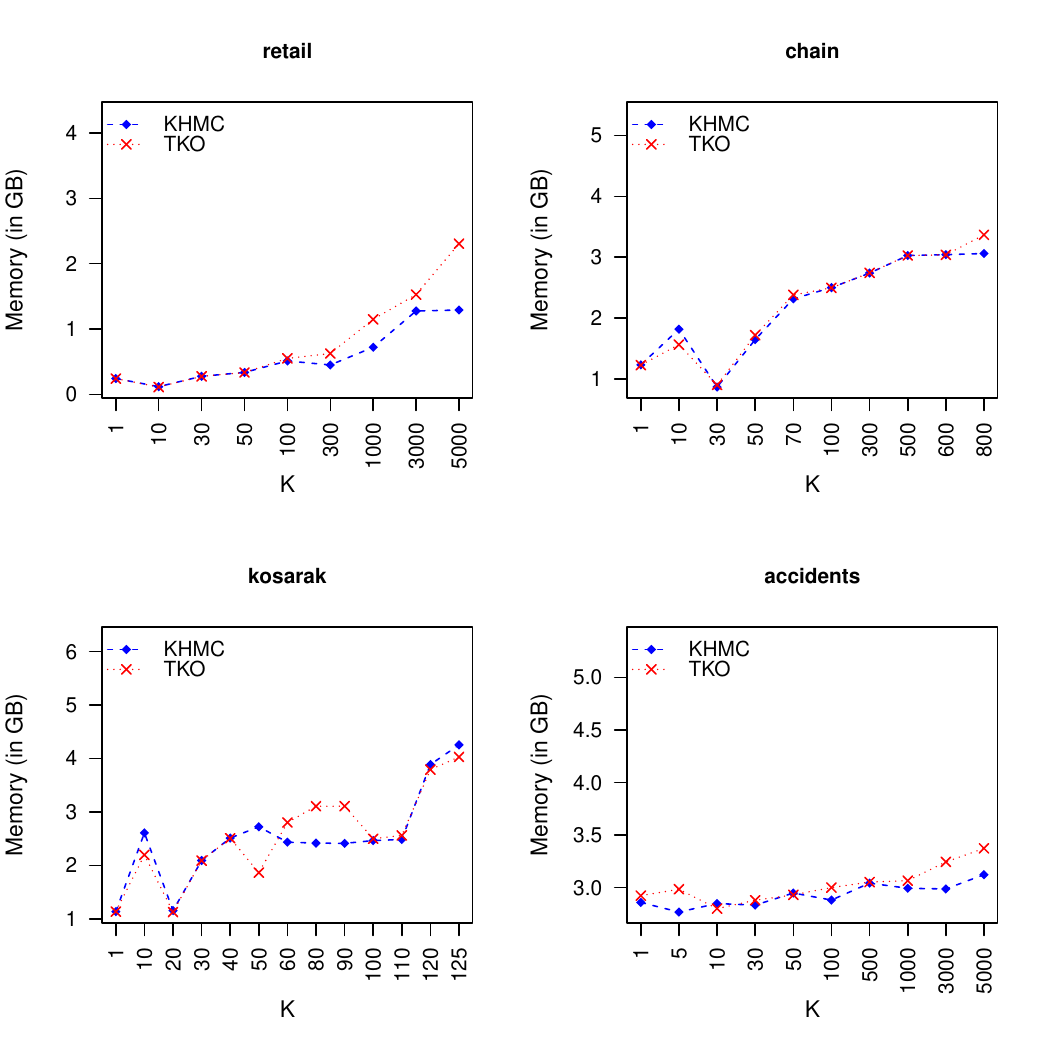}
	\caption{Memory consumption performance on sparse datasets}
	\label{fig:memorySparse}
\end{figure}

The results of memory consumption performance of these algorithms are given in Figure~\ref{fig:memorySparse}. One can observe from the results that there is no significant difference in memory consumption performance of both these algorithms on sparse datasets. 

\begin{figure}[!h]
	\includegraphics[height=8cm,width=11cm,page=4]{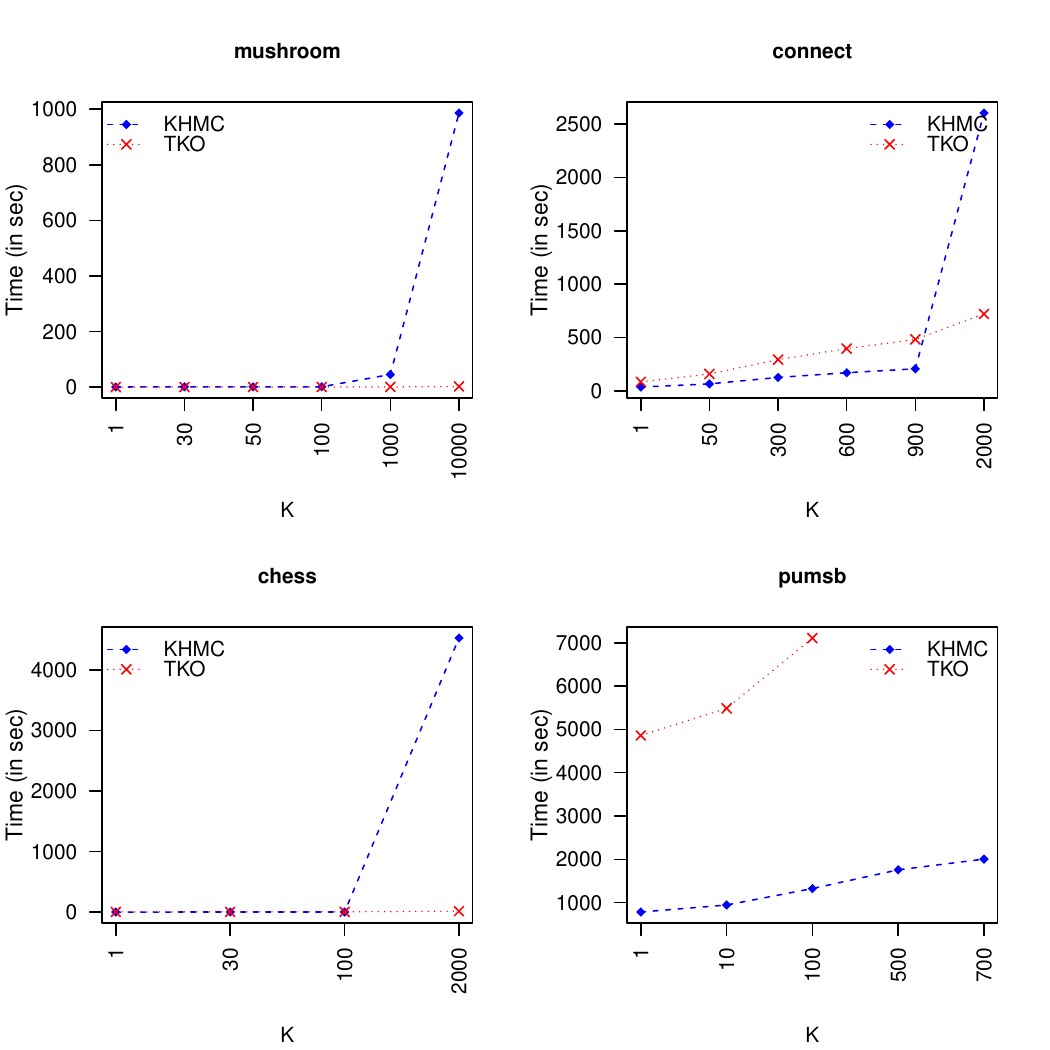}
	\caption{Runtime analysis of KHMC and TKO on dense datasets}
	\label{fig:runtimeDense}
\end{figure}

\begin{figure}[!h]
	\includegraphics[height=8cm,width=11cm,page=5]{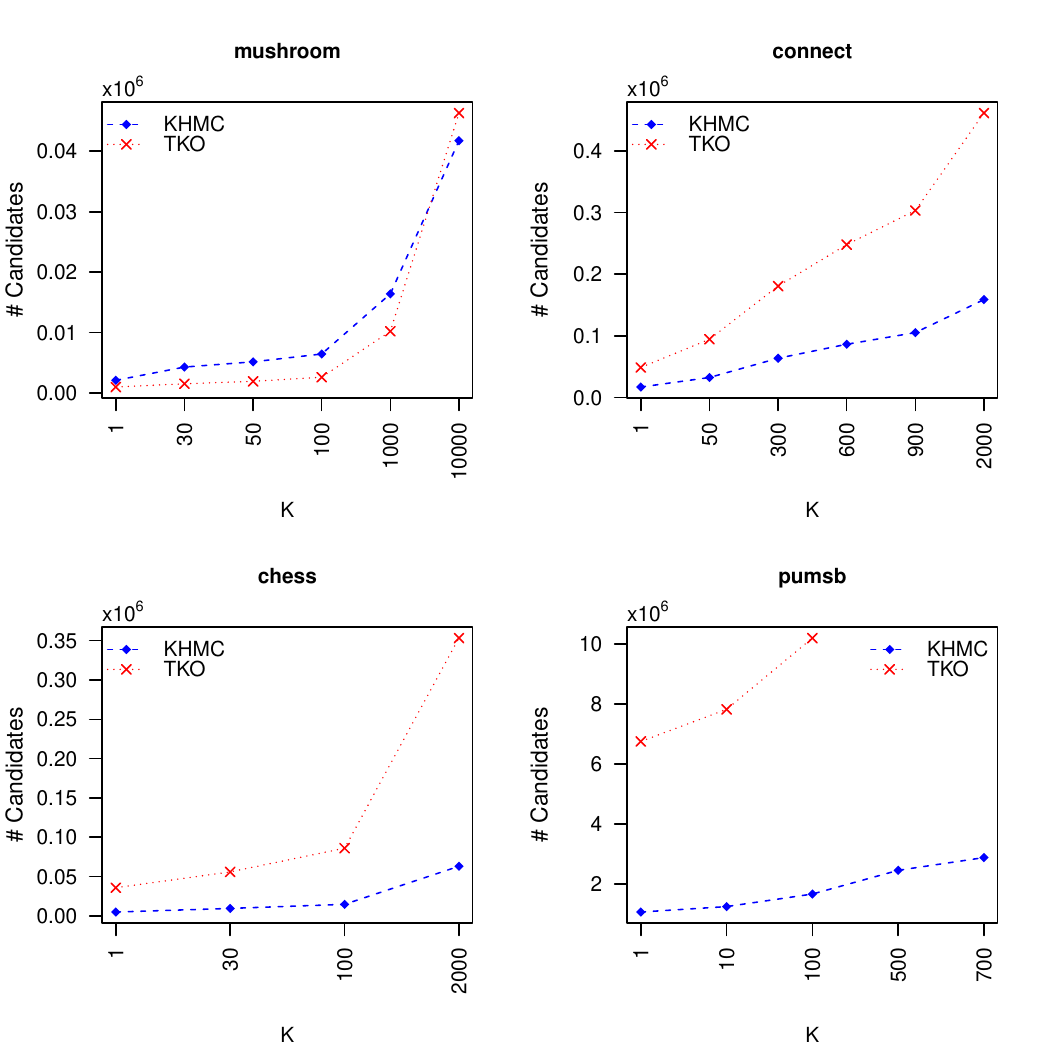}
	\caption{Number of candidates generated on dense datasets}
	\label{fig:candidatesDense}
\end{figure}

\begin{figure}[!h]
	\includegraphics[height=8cm,width=11cm,page=6]{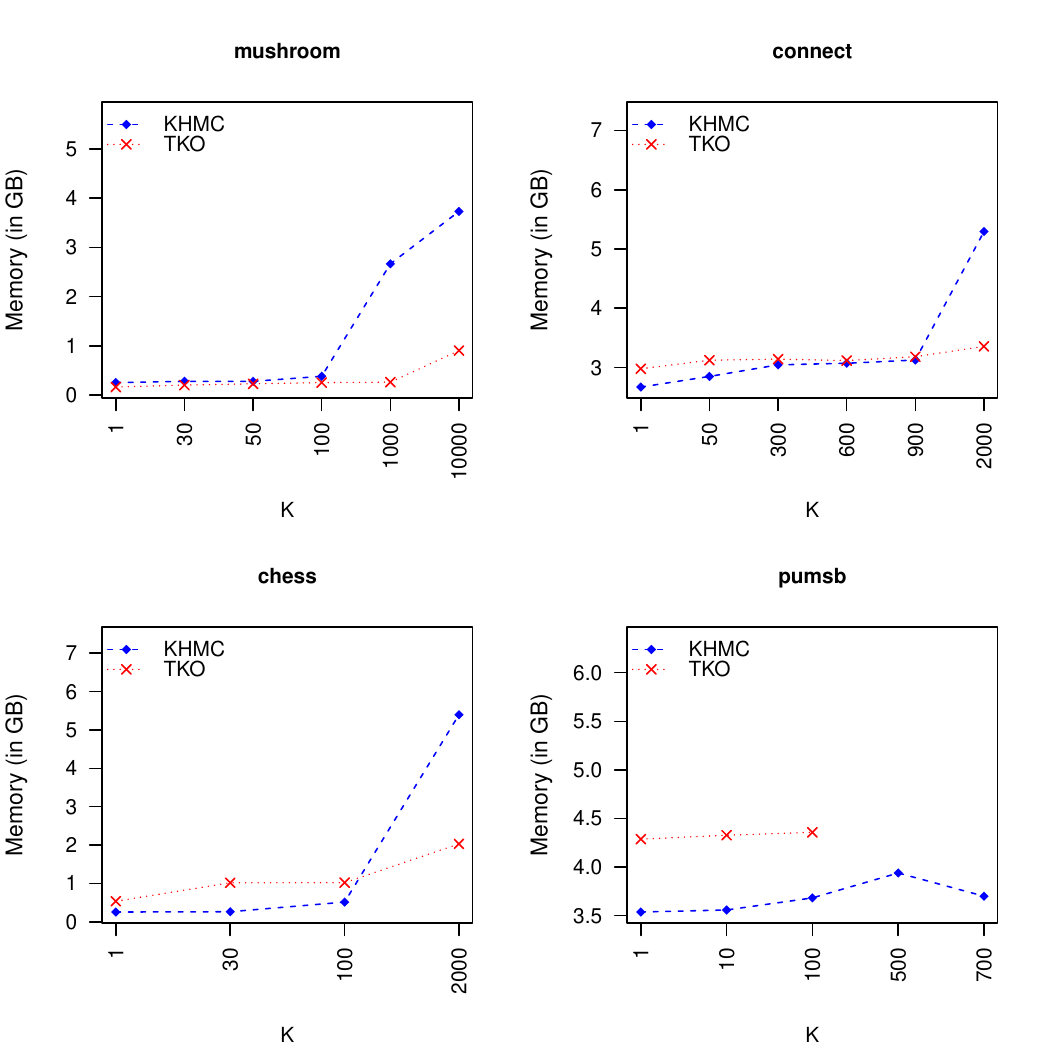}
	\caption{Memory consumption performance on dense datasets}
	\label{fig:memoryDense}
\end{figure}

In the next set of experiments, we analyzed the performance of the two algorithms on dense benchmark datasets. The results of our experiments are shown in Figures~\ref{fig:runtimeDense}, \ref{fig:candidatesDense} and \ref{fig:memoryDense}. The runtime performance analysis results reveal that on three out of four dense benchmark datasets, the TKO algorithm works significantly better. At higher values of $K$, the performance of KHMC algorithm degrades significantly. We also observed that KHMC algorithm runs out of memory at higher values of $K$. The degradation in performance of KHMC can be attributed to the use of coverage based threshold raising strategy (COV). The COV strategy requires estimating utilities of supersets based on coverage of single items. As the average length of transactions are longer in the case of dense datasets, the coverage based threshold raising strategy is quite expensive. For instance, in the case of mushroom dataset, the total number of candidates evaluated by both of these methods are almost similar (refer to Figure~\ref{fig:candidatesDense}). But, the runtime performance of KHMC was observed to be poor. The poor performance of KHMC can be attributed to the expensive coverage evaluation process. 

Our analysis of candidate sizes generated by TKO and KHMC on dense datasets reveal interesting insights. In almost all of the datasets studied, the number of candidates evaluated by TKO is much higher. This can be attributed to two reasons: (1) KHMC algorithm uses an EA strategy to abandon unpromising candidates early, and (2) TKO algorithm uses a EPB strategy to reorder the candidate itemsets based on their estimated utilities (refer to property~\ref{property:epb}). It is to be noted that small changes in ordering of candidates can significantly impact the performance of algorithms, especially when the average length of transactions (and hence itemsets) are longer. In the case of pumsb dataset, the average length of transaction is 74. We conjecture that the EPB strategy is likely to perform poorly for very long and dense datasets. Further research is required to validate this conjecture and also understand the performance of each of the individual pruning strategies adopted by these methods. 

The memory consumption performance results of each of these algorithms show marginally better results for the TKO algorithm. This can be attributed to the space requirement for the COV strategy that requires evaluation of superset items based coverage of single items. As the size of the supsersets to be evaluated are likely to higher for dense datasets, the memory consumption requirement tend to be much higher. In the case of pumsb dataset, the memory requirement of TKO algorithm was found to be marginally higher than the KHMC algorithm.  

Overall, we observe that the KHMC algorithm was found to work well on sparse benchmark datasets. On the other hand, the TKO algorithm performs better on most of the dense benchmark datasets studied. Our analysis reveals that COV strategy used in KHMC algorithm performs very poorly for large and dense datasets. Furthermore, the EPB strategy adopted in TKO algorithm was found to be quite expensive for datasets with very long transactions (e.g., pumsb dataset). Future research can explore the trade-offs involved in individual threshold raising strategies (PE, NU, MD, MC, SE, PUD, RIU, RSD, SEP, RUC, CUD and COV) and pruning properties (TWDC, DGU, DGN, DLU, DLN, RUZ, EPB, U-Prune, EUCS, EA and TEP).

\section{Top-K high utility pattern mining variants}
\label{sec:5}

A few extensions to the basic top-k HUI mining problem have been explored in the recent literature. We briefly review each of these methods in this section.
\par 
Yin et al \cite{yin2013efficiently} present a TUS algorithm for mining top-k high utility sequential patterns. The TUS algorithm uses two new threshold raising strategies (pre-insertion and sorting) and one pruning strategy (sequence reduced utility) to filter unpromising candidates and improve the performance of mining. The authors demonstrate the superiority of their method against a baseline top-k high utility sequential pattern method (TUSNaive). 
\par 
Zihayat et al \cite{zihayat2014mining} propose T-HUDS method for determining top-k high utility patterns over data streams. The method uses a compressed tree data structure, that is similar to UP-Tree, named HUDS-tree. It also uses a new utility estimation method (PrefixUtil) to prune the search space and efficiently mine top-k HUIs. The T-HUDS method works in two-phases. In the first phase, the HUDS-tree is constructed and mined to generate a set of potential top-k HUIs. Subsequently, in the second phase, the actual top-k HUIs are identified by computing the exact utilities of potential top-k HUIs. 
\par One of the more recent works on top-k HUI mining over streams is by Dawar et al \cite{dawar2017miningStreamingTopK}. The authors present a one-phase approach to efficiently mine top-k HUIs over data streams without generating intermediate candidates as in T-HUDS \cite{zihayat2014mining}. The authors compare their method against the two-phase T-HUDS method and demonstrate its usefulness on both spare and dense benchmark datasets. 
\par 
Dam et al \cite{dam2017efficientOnShelfTopK} present a top-k on-shelf high utility pattern mining method named KOSHU. Their method considers items with either positive or negative unit profits. The KOSHU algorithm scans the database twice to construct the 1-itemset utility list. The generated 1-itemset utility list is then used to explore the search space and mine all the on-shelf top-k high utility patterns. KOSHU uses three pruning strategies and two threshold raising strategies to effectively mine on-shelf high utility patterns. The new pruning strategies used in KOSHU include: Estimated Maximum Period Rate Pruning (EMPRP), Period Utility Pruning (PUP) and Concurrence Existing of a pair 2-itemset Pruning (CE2P). The threshold raising strategies used in KOSHU include: (1) Real 1-Itemset Relative Utility (RIRU), that is inspired by the RIU strategy (refer to strategy~\ref{strategy:riu}), and (2) Real 2-Itemset Relative Utility (RIRU2). The authors conduct rigorous experiments on real and synthetic datasets to show the utility of the KOSHU method.

\section{Open issues and future research opportunities}
\label{sec:6}
High utility itemset mining is one of the very active research areas in data mining. Numerous algorithms have been proposed in the last decade for mining basic high utility itemsets. The top-k HUI mining aims to address some of the core limitations of basic HUI mining. Some of the key algorithms proposed in the literature have been extensively reviewed in the foregoing sections. In this section, we outline key issues in current top-k HUI mining methods and discuss future research opportunities. 
\par 
\textit{Nature of profitability of items.} Almost all of the current top-k HUI mining methods support only the positive unit profits, except KOSHU \cite{dam2017efficientOnShelfTopK} that considers both positive and negative unit profit items. In addition, it is possible for the same item to take on positive or negative unit profits at different points in time. Future work could consider the support for negative unit profit items \cite{fhnKBS2016} as well as the mix of positive and negative unit profit items at individual transaction levels. These extensions are non-trivial and require design of newer threshold raising and pruning strategies to efficiently mine top-k HUIs.
\par 
\textit{Impact assessment of threshold raising strategies.}
Numerous threshold raising strategies have been introduced in the literature. In this survey paper, we have identified twelve different threshold raising strategies and made a qualitative comparison of different approaches. It would be interesting to conduct rigorous performance analysis of different threshold raising strategies on benchmark datasets and assess the trade-offs involved. Research in this direction would be useful to discover new threshold raising strategies to advance the field further.
\par 
\textit{Design of pruning properties.}
The review paper discussed several pruning properties adopted in the literature on top-k HUI mining. Almost all of the pruning properties are direct application of properties designed for basic HUI mining. Future work can explore design of new pruning properties to significantly improve the performance of top-k HUI mining. 
\par 
\textit{Adopting advances in basic HUI mining.}
Several advances have been made in the basic HUI mining literature in the last few years. For example, EFIM \cite{efimKBS2017} explores a database projection method to significantly improve the performance of HUI mining. HMiner \cite{krishnamoorthy2017hminer} is another more recent HUI mining method that uses a compressed utility list data structure for efficiently mining HUIs. These methods have been proven to be several orders of magnitude faster compared to other state-of-the-art methods in the literature. It would be interesting to extend these ideas in the context of top-k HUI mining and substantially improve the performance of top-k HUI mining. 
\par 
\textit{Explore Top-K HUI variants}
There are very few research works on top-k HUI mining variants in the literature. Some of the top-k HUI variants studied in the literature include: on-shelf utility mining, sequential pattern mining and data stream mining. Future work could consider more algorithmic improvements on these top-k variants. It would also be interesting to study other HUI mining variants such as imprecise and uncertain HUIs \cite{gan2017miningUncertain,lin2017efficientlyUncertain}, and high average utility itemsets \cite{lin2016avgHUImmu,lin2016efficient}.

\section{Conclusions}
\label{sec:7}

This paper systematically analyzed the top-k HUI mining methods in the literature. It reviewed and compared different one-phase and two-phase methods in the literature. The key data structures, threshold raising strategies and pruning properties used in the top-k HUI mining methods were discussed in detail. A performance evaluation of the state-of-the-art methods (TKO and KHMC) were also made. Our results reveal that the KHMC \cite{duong2016efficientKHMC} method offers the best performance on sparse benchmark datasets. The TKO \cite{tseng2016efficientTKO} method was found to work well for most of the dense benchmark datasets studied. 
\par The top-k HUI mining problem variants such as on-shelf mining, data stream mining and sequential pattern mining were also analyzed. Furthermore, the paper outlined future research opportunities in the area of top-k HUI mining. This survey paper is likely to be beneficial for researchers to explore and understand the developments in the field of top-k HUI mining, assess the key gaps in the literature and advance the state-of-the-art in top-k HUI mining.

\bibliography{biblio}

\end{document}